# Extraordinary Lifetime Enhancement of Coherent Phonon-Amplitude Modes in the Excitonic Insulator Phase of $Ta_2Pd_3Te_5$

**Anjan Kumar N M[1*], Shuhan Wang[1], Snehashish Chatterjee[2], Yan Zhu[1], MinJae Kim[1,3], Tobias Ritschel[1], Elaheh Sadrollahi[1], Jochen Geck[1,5], Achim Rosch[4], Chandra Shekhar[2], Claudia Felser[2,5], Stefan Kaiser[1,5*]**

[1] Institute of Solid State and Materials Physics, Technische Universität, Dresden, Germany.

[2] Topological Quantum Chemistry Group, Max Planck Institute for Chemical Physics of Solids, Dresden, Germany

[3] Korea Institute of Science and Technology, Seoul, South Korea.

[4] Institute for Theoretical Physics, University of Cologne, Cologne, Germany.

[5] Würzburg-Dresden Cluster of Excellence ctd.qmat, Technische Universität Dresden, 01062 Dresden, Germany

*Email: anjan.naralapura@tu-dresden.de ; stefan.kaiser@tu-dresden.de

**Abstract:** The excitonic insulator (EI) is an electronic phase of condensed excitons. However, in many prototypical materials the presence of a concurrent structural transition complicates the identification of a purely electronic origin of the ordered state. Here, we investigate $Ta_2Pd_3Te_5$, in which an EI phase develops below $T_C \approx 100$ K in the absence of a detectable structural transition. We characterize the excitonic condensate via its coherent phonon-amplitude response. Most strikingly, an extraordinarily strong lifetime enhancement of these modes sets in below $T_C$ which we establish as a new and robust fingerprint linked to exciton condensation. We discuss possibilities to capture the enhancement when taking into account a coupling of excitonic and lattice effects in line with the coupled phonon-amplitude response. Within an order-parameter polaron picture coupling to the excitonic state may dress the phonon mode and thereby suppress its relaxation and a phenomenological model combines the anharmonic phonon response and a reduced electronic scattering due to the gap opening in the EI phase.

**Introduction:**

An excitonic insulator (EI) is an interaction-driven condensate of bound electron–hole pairs (excitons). It can emerge for high exciton binding energies exceeding either the bandgap in a small bandgap semiconductor or the small bandwidth in a semimetal [1-5]. The transition can be described as a Bose–Einstein–like condensation (BEC) of preformed excitons or as Bardeen–Cooper–Schrieffer–type (BCS) instability at the formation of excitons, respectively [6]. The BEC regime favours localized excitons with large binding energies, whereas in the weak-coupling BCS regime the exciton wavefunctions overlap. In the first case the high binding energy results in short exciton lifetimes that prevent stable phases of preformed excitons. In the second case the low exciton binding energies are in competition with their requirement to exceed the bandwidth of the semimetal. Whether the excitonic insulator can be experimentally realized in a real material remains an open question. Promising candidate materials are 1T-$TiSe_2$ and $Ta_2NiSe_5$, both of which have been extensively investigated experimentally and theoretically [7-17]. A central difficulty is that in these systems the transition into the EI state concurs with a structural phase transition. In 1T-$TiSe_2$ an indirect gap results in an insulating CDW ground state [ 11, 18-20]. $Ta_2NiSe_5$ is a direct gap semiconductor but at the semiconductor-insulator transition also a structural phase transition occurs [7, 17]. In both systems a debate if these transitions are structural or electronic in nature and about fingerprints for an EI state is ongoing. Among others time-resolved methods have contributed to this question [10-12, 20-30].

In particular time-resolved measurements allowed characterizing the coherent response of the EI ground state upon photoexcitation by probing the coherent amplitude [27, 29-31] and phase modes [29]. Pump-probe experiments on $Ta_2NiSe_5$ have revealed coherent collective oscillations interpreted as coupled phonon-amplitude modes of the condensate [27]: Their amplitude tracks the excitonic order parameter as a function of temperature and excitation fluence. The latter exhibits a characteristic depletion behaviour following the melting of the condensate for high excitation fluences. The coupling of the phonon to the excitonic order parameter can be experimentally verified by a nonlinear frequency shift of the collective modes [27] and be described within exciton-phonon interaction picture [30]. The coupling to the lattice is also emphasized in the Fano-shaped responses in Raman [32-34], infrared [35] and THz experiments [36].

In the context of a dominating electronic nature of the EI formation a new candidate system, $Ta_2Pd_3Te_5$, has gained attention, where exciton condensation is discussed without a concomitant structural symmetry breaking [37-43]. Here, we employ time-resolved probes on this novel system in analogy to the successful characterization in $Ta_2NiSe_5$ [27]. Importantly, we identify coupled phonon-amplitude modes despite the absence of a detectable structural phase transition in $Ta_2Pd_3Te_5$. Beyond that, we find a striking novel experimental hallmark linked to exciton condensation: In the EI phase the lifetime of the coherent mode enhances beyond what is typically expected for usual phonon anharmonicities. We discuss different models including the influence of the excitonic gap opening, and the stabilization of phonons/excitons within polaronic complexes in the EI phase and use these to establish the lifetime enhancement as a new fingerprint of the excitonic condensate.

**Results:**

$Ta_2Pd_3Te_5$ is a van der Waals material with a quasi-one-dimensional structure as shown in Fig. 1(a). Transport measurements report a metal-insulator/semiconductor transition near $T_C^* \approx 365$ K and the opening of a gap of up to 55 meV at low temperatures from ARPES measurements [37] that is confirmed to be of excitonic nature in Ref. [41]. However, scanning tunneling spectroscopy observes the opening of a gap only below $T_C \approx 100$ K [39], which grows to about $2\Delta \approx 80$ meV at 5 K. Raman spectroscopy identified a potential excitonic mode in the electronic background that shows strong renormalization around $T_C \approx 100$ K [44]. X-ray diffraction, Raman spectroscopy of the phonon modes, and scanning tunneling microscopy find no evidence for charge-density-wave order or lattice symmetry breaking at any of these temperatures [37].

Here, we perform non-degenerate pump-probe reflectivity measurements (supplementary materials, Sec. 1) on high-quality single crystals of $Ta_2Pd_3Te_5$ (supplementary materials Sec. 2). The time-dependent reflectivity changes in Fig. 1(b) describe the electronic and phononic response of the system upon photoexcitation at 20 K. The electronic signal shows a double exponential decay that we can capture with the usual expression

$$\frac{\Delta R(t)}{R} = \left[A_1 e^{-\frac{(t-t_C)}{\tau_1}} + A_2 e^{-\frac{(t-t_C)}{\tau_2}} + B\right] \times \left[\mathrm{erf}\left(\frac{(t-t_C)}{t_r}\right) + 1\right] \qquad Eq.\ 1$$

where the error function term accounts for the finite rise time ($t_r$) of the onset. $A_1$ and $A_2$ are the coefficients of the fast and slow relaxation channels, respectively, with $\tau_1$ representing the fast decay relaxation time and $\tau_2$ corresponding to the slower relaxation component and a constant parameter $B$. The data fitting and the detailed evolution of the extracted fit parameters is provided in the supplementary materials Sec. 3.

The striking feature of the signal is long-lived coherent oscillations that extend up to hundreds of picoseconds (inset of Fig. 1(b)) that will be discussed further below. In the electronic response the fast relaxation component, with a characteristic timescale of $\tau_1 \approx 1$ ps, exhibits only a weak anomaly near $T_C$, consistent with carrier thermalization dominated by electron–electron scattering (see supplementary materials, Sec. 3). In contrast, the slower electron–phonon relaxation component $\tau_2$ (Fig. 1(c), top panel) displays an activated temperature dependence, peaking near $T_C$ ≈100 K, with characteristic timescales extending to a few hundred picoseconds. This dynamical energy transfer is in full agreement with the activation energy inferred from our transport measurements indicating a thermodynamic phase transition. Also, the amplitude of the slow component $A_2$ shows a very sharp change at 100 K as shown in Fig. 1(c) (bottom panel). This sets a characteristic temperature scale for our response similar to the one reported in the tunnelling experiment [39]. Notably, our single-crystal X-ray diffraction measurements (supplementary materials Sec. 2) reveal no evidence of a structural phase transition between the measured temperatures of 290 K and 10 K. At both temperatures, $Ta_2Pd_3Te_5$ crystallizes in the orthorhombic structure with space group Pnma (No. 62) as shown in Fig. 1(a).

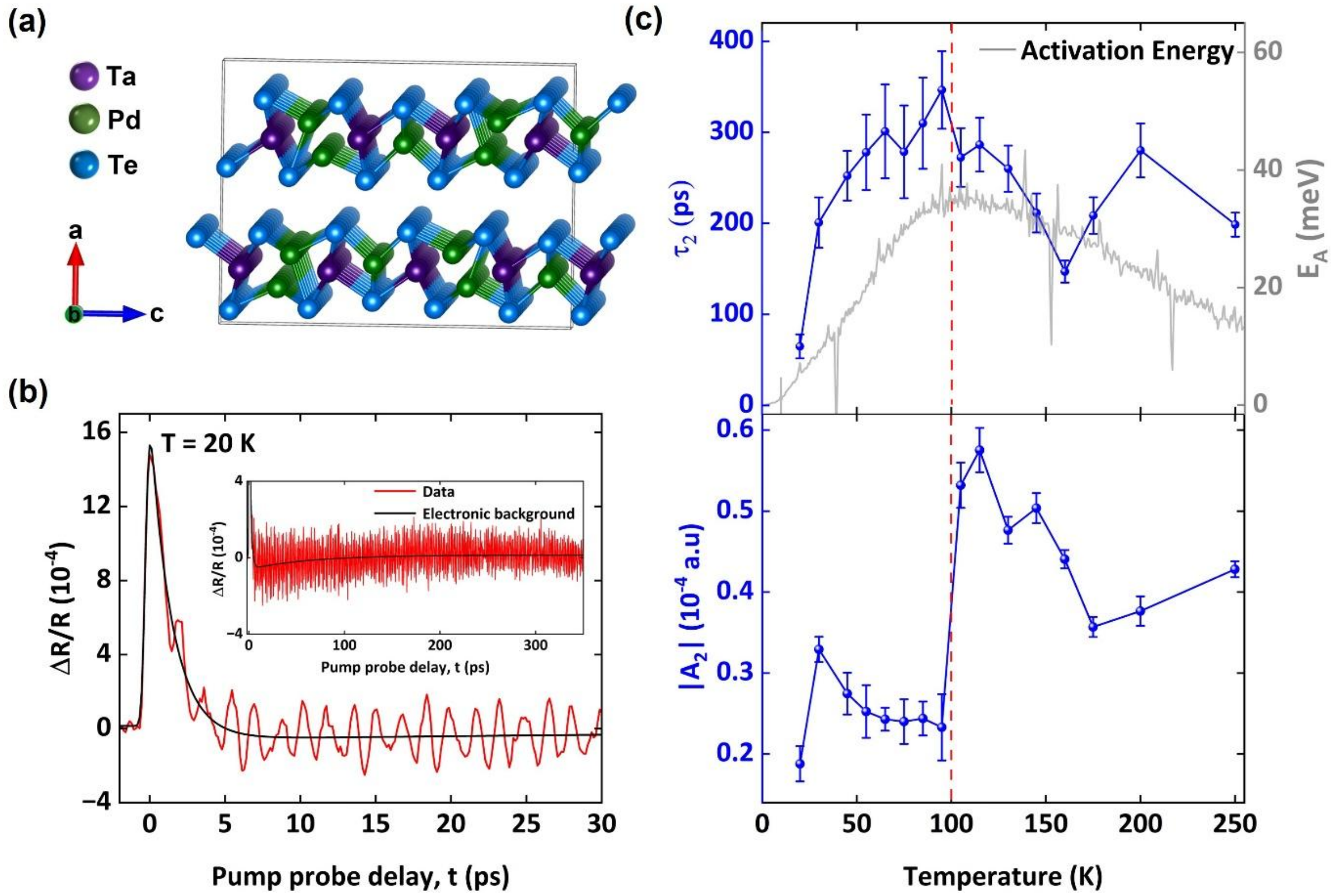

Figure 1: **Structural and ultrafast optical characterization of $Ta_2Pd_3Te_5$.** (a) Crystal structure of $Ta_2Pd_3Te_5$, in which each unit cell contains two $Ta_2Pd_3Te_5$ layers stacked along the a-axis and weakly coupled by van der Waals interactions. Within each layer, Ta (purple) and Pd (green) atoms form one-dimensional chains along the b-axis, coordinated by Te (blue) atoms. (b) Photoinduced reflectivity changes ($\Delta R/R$) at 20 K and a pump fluence of 20 μJ/cm$^2$ as a function of pump–probe delay time $t$. The inset shows the long-lived oscillations. The experimental data (red solid line) are fit with a bi-exponential function (black solid line) described in Eq. (1). (c) Extracted temperature dependence of relaxation time $\tau_2$ (top panel) and the corresponding amplitude $A_2$(bottom panel). The solid gray line in the top panel denotes the activation energy obtained from temperature-dependent transport measurements (supplementary materials Sec. 2), peaking at the transition temperature, $T_c \approx 100$ K (red dashed line).

For the coherent phononic response we subtract the electronic background. Figure 2(a) shows the temperature dependence of the remarkably long-lived oscillations. The time-domain data shows that for high temperatures the oscillations are damped resulting in typical phonon lifetimes of up to 100 ps. However, below $T_c$ ≈ 100 K no obvious damping of the oscillations in the time-domain is visible within our measurement range. A Fourier analysis in Fig. 2(b) reveals impulsively stimulated $A_g$ modes at 0.616 THz, 0.85 THz, and 1 THz (not shown) [44, 45]. Since the 1 THz mode is comparatively weak, we focus on the two dominant low-frequency modes. Figure 2(a) shows 0.55-0.9 THz band-pass filtered time-domain oscillations that we fit with an exponentially damped bi-cosine function (supplementary materials, Sec. 3):

$$\left(\frac{\Delta R(t)}{R}\right)_{Osc} = A_1^{ph} e^{-\left[\frac{t}{\tau_1^{ph}}\right]} \cos\left(2\pi\nu_1^{ph} t + \phi_1^{ph}\right) + A_2^{ph} e^{-\left[\frac{t}{\tau_2^{ph}}\right]} \cos\left(2\pi\nu_2^{ph} t + \phi_2^{ph}\right) \qquad Eq.2$$

where $A_i^{ph}$ 's are the overall amplitudes, $\tau_i^{ph}$ 's are the phonon lifetimes, $\nu_i^{ph}$ 's are the oscillation frequencies and $\phi_i^{ph}$ 's are the phases.

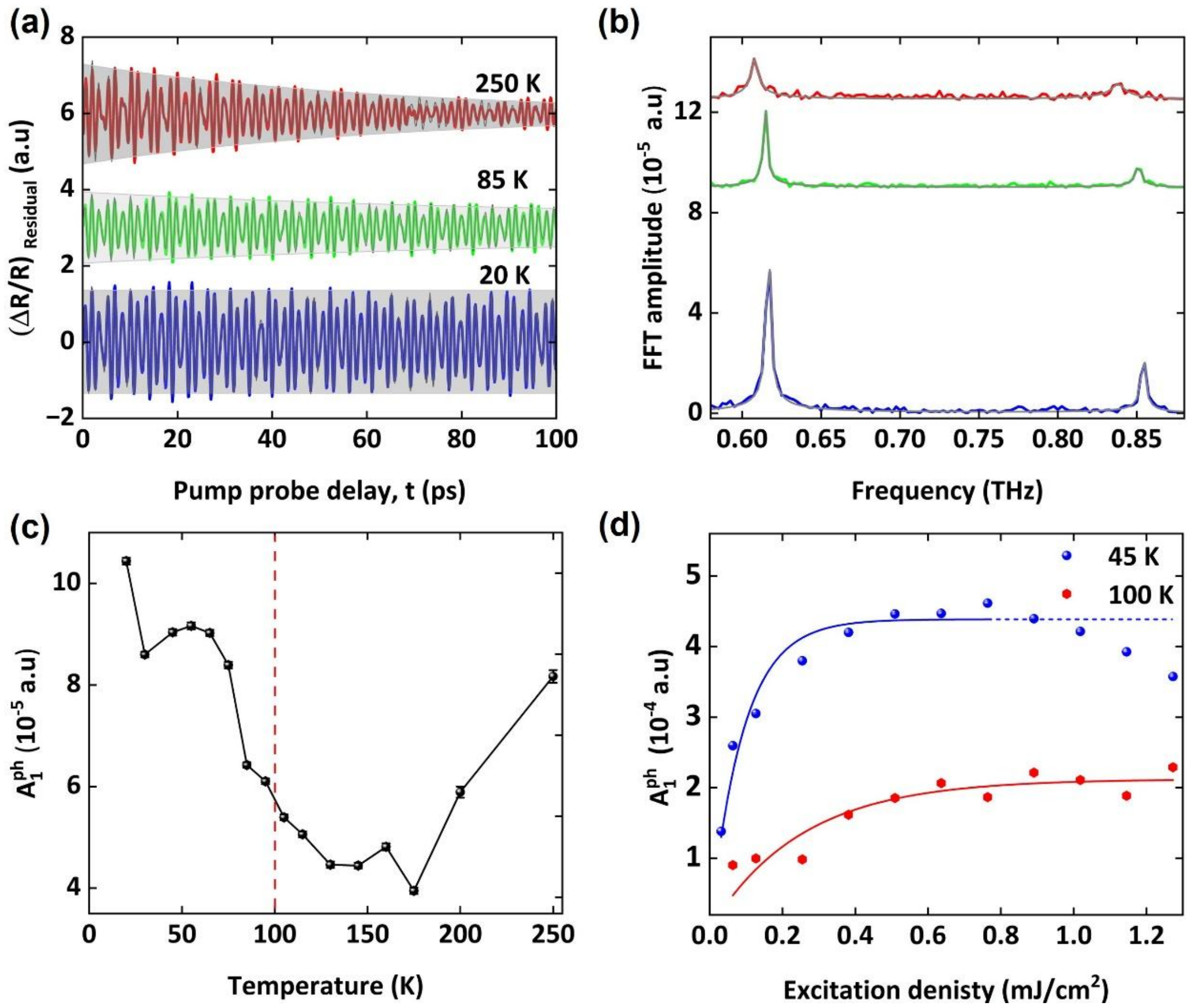

Figure 2: **Temperature and fluence-dependent dynamics of coherent phonon- amplitude modes in $Ta_2Pd_3Te_5$.** (a) Oscillatory components extracted from the electronic background of the $(\Delta R/R)$signals at a pump fluence of 20 μJ/cm$^2$ at selected temperatures and their fits (solid gray lines) according to Eq (2). The shaded regions highlight the damped/non-damped behaviour of the responses. (b) Fast Fourier transform (FFT) spectra of these oscillations and their fits (solid gray lines) shown in (a). Traces in (a) and (b) are vertically offset. (c) Temperature dependence of the amplitude of the 0.61 THz mode, showing a pronounced enhancement below the transition temperature, $T_c = 100\ K$ (dashed line). (d) Fluence dependence of the 0.61 THz mode amplitude measured at 45 K and 100 K. The solid lines represent fits to a saturation model, $Aexp\ (-\rho/\rho_0)$. The 45K data show a depletion effect for densities beyond 0.8 mJ/cm$^2$.

In the following we are going to discuss the 0.61 THz mode with the larger oscillation amplitude. The qualitatively similar response of the 0.85 THz mode is shown in the supplementary materials Sec 4. As shown in Fig. 2(c) the amplitude decreases on cooling down to 200 K. Then the amplitude remains fairly constant until 120 K before it increases around the transition temperature, $T_c$. The enhancement of the amplitude below $T_c$ qualitatively tracks the temperature dependence of the gap extracted from ARPES measurements [39]. In analogy to $Ta_2NiSe_5$ this could be interpreted as the response of a coupled phonon–amplitude mode tracing the coherent order parameter in the EI phase [27, 30]. We note that the response in $Ta_2Pd_3Te_5$ does not exhibit the well-defined mean-field-like order-parameter behaviour as previously reported for $Ta_2NiSe_5$. The more complex temperature-dependence (including the peak structure around 55 K) goes beyond the focus of this study and is likely due to the presence of in-gap states as reported in Ref. [39]. To foster our interpretation of a coupled mode tracing the order parameter Fig. 2(d) shows the fluence dependence of the mode at 100 K (red) and 45 K (blue). At high temperatures the mode behaves like a coherent phonon where the oscillation amplitude increases with the excitation fluence until saturation sets in [46, 47]. In contrast, in the EI phase, below $T_c$, the mode first shows a similar saturation behaviour around fluences of 0.2 mJ/cm$^2$ followed by a clear reduction of the signal at fluences beyond 0.8 mJ/cm$^2$. The latter is due to a depletion of the EI condensate as explained for the corresponding phonon-amplitude response reported in $Ta_2NiSe_5$ [27].

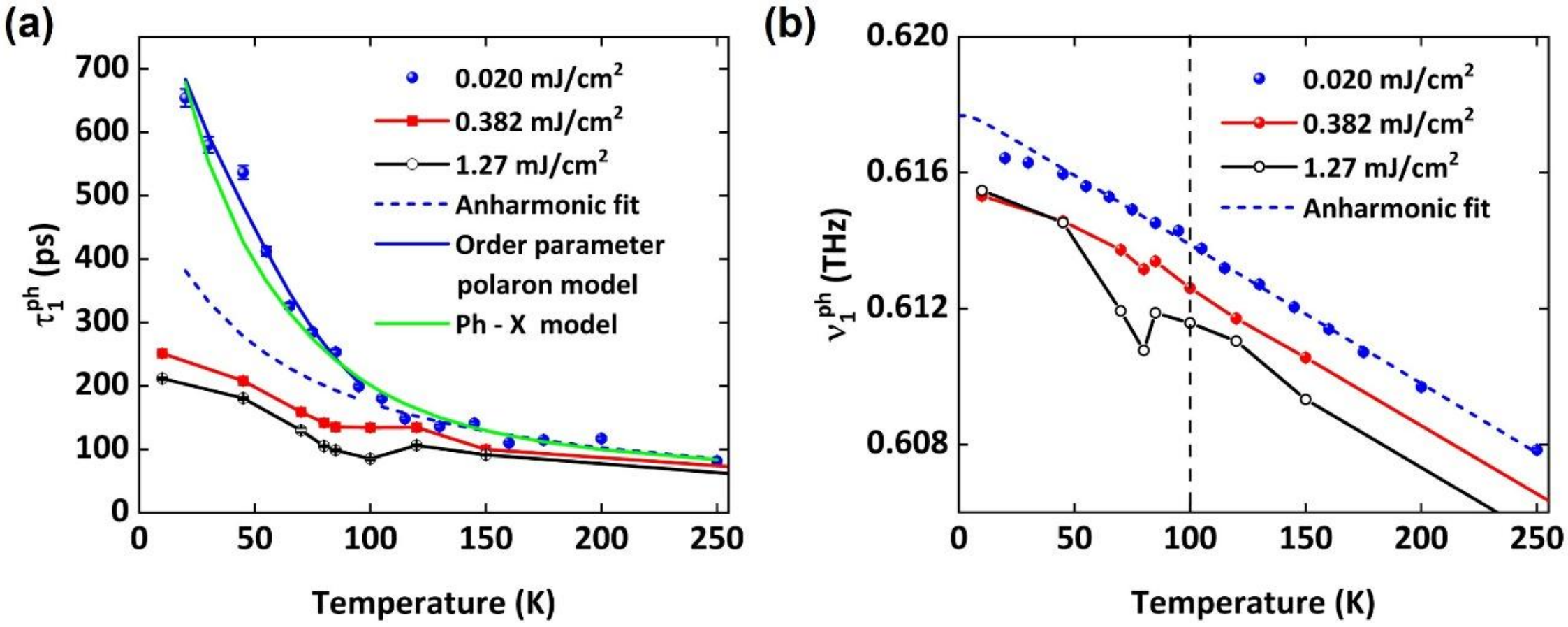

Figure 3: **Lifetime enhancement of 0.61 THz coherent phonon- amplitude mode in the EI phase of $Ta_2Pd_3Te_5$** (a) Temperature dependence of the extracted lifetime of the 0.61 THz mode at selected excitation fluences. The lifetime below 100 K shows clear deviation from an anharmonic lattice model (blue dashed lines). The lifetime below 100 K is captured by an order parameter polaron model (blue solid line) and the coupled phonon- exciton (Ph-X) model (green solid line) (b) Temperature dependence of the frequency of the 0.61 THz mode at representative excitation fluences. The low fluence case the mode frequency is reproduced by the anharmonic model above $T_C$, whereas below $T_C$, it deviates from the anharmonic trend (blue dashed line). At higher fluences, a dip-like feature emerges below the $T_C$.

While this coupled response gives us a first hint on the formation of an EI state we now turn our attention to a novel experimental fingerprint for the condensation of excitons. While coherent phonon oscillations at high temperatures exhibit finite lifetimes, the coherent phonon-amplitude oscillations below $T_C$ do not show signs of damping resulting in extraordinarily long lifetimes of the modes. The corresponding temperature dependence is shown in Fig. 3(a) for the 0.61 THz mode (see supplementary materials Sec.4 for the 0.85 THz mode). For a low excitation density of 20 µJ/cm$^2$ (blue) we find a linear increase of the phonon lifetime on cooling to the transition temperature $T_C \approx 100$ K. This behaviour and lifetimes of up to 150 ps are typical for coherent phonons [48]. However, cooling below 100 K a highly non-linear increase of the modes lifetime sets in, reaching unprecedented ~ 654 ps at 20 K (a value currently limited only by our experimental time window, supplementary materials Sec. 3) for a coherent phonon in a solid-state system. For comparison, the longest reported coherent phonon lifetime so far is ≈ 211 ps at 5 K in ZnO [49]; the lifetime observed here in $Ta_2Pd_3Te_5$ represents a three-fold enhancement over this benchmark. Thus, it is unlikely that this non-linear lifetime enhancement can be understood as conventional lifetime of a coherent phonon. In the discussion we show that it goes beyond the typical enhancement based on phonon anharmonicities (see also supplementary materials Sec. 5). We discuss two models that can capture the lifetime enhancement: an exciton–polaron coupling in the condensate phase, and as a phenomenological model of an anharmonic phonon decay with a reduced electronic scattering rate in the condensate. Similar stabilization of states in condensates was anticipated in photoluminescence studies for excitons in $NiPS_3$ [50] or ferromagnetic resonance linewidth narrowing for magnons in YIG [51]. The lifetime enhancement in the condensate crucially depends on the excitation fluence. Increasing it to above 0.2 mJ/cm$^2$ the enhancement effect is heavily suppressed (Fig. 3(a), red). This fluence characterizes the onset of the saturation regime of the amplitude oscillations seen in Fig. 2(d). The complete fluence dependent of lifetime is discussed in the supplementary sec. 6. Increasing the fluence into the depletion regime at a fluence of 1.27 mJ/cm$^2$ the lifetime enhancement fully disappears (Fig. 3(a), black). In the latter case the temperature dependence of the lifetime shows a dip-like feature around $T_C$ that together with a corresponding feature in the temperature dependence of the frequencies (Fig. 3(b)). These may hint at a light-induced structural transition in the high fluence regime (supplementary materials Sec. 6)

**Discussion**

Now, we discuss the possible origin of the remarkable lifetime enhancement of the coherent mode below the transition temperature. Raman measurements on $Ta_2Pd_3Te_5$ show that the temperature dependence of the linewidths and frequencies of the ($A_g$) phonon modes is well described by an anharmonic model, in which an optical phonon decays into two acoustic phonons with equal and opposite momenta [44]. Here, above the transition temperature, $T_C$, the measured phonon lifetime is captured by the anharmonic decay model (Fig. 3(a), and supplementary materials Sec 5). Upon cooling below $T_C$, however, the experimentally observed lifetime shows a pronounced enhancement relative to the low-temperature extrapolation of the anharmonic model. The mode frequency is also well described by the anharmonic model above $T_C$, whereas below $T_C$, a slight deviation is found (Fig. 3(b)). This indicates that the observed lifetime enhancement cannot be explained solely by lattice anharmonicity. One possible contribution is the suppression of decay channels into low-energy electronic excitations. Upon entering the condensate state, the opening of the excitonic gap suppresses these channels, resulting in a narrowing of the phonon linewidth and a corresponding increase in the phonon lifetime. To examine this scenario we modeled the inverse phonon lifetime using a modified Arrhenius expression and a phenomenological decay model incorporating radiative and non-radiative recombination channels (see supplementary materials Sec. 5b). These analyses yield activation energy and temperature-dependent gap values of only a few meV, substantially smaller than the excitonic gap of approximately 80 meV at 5 K reported in the literature. This discrepancy suggests that the suppression of electronic damping alone cannot fully account for the exceptional lifetime enhancement.

We have seen that the amplitudes of the coherent oscillations below the critical temperature can be understood in the framework of a coupled phonon-amplitude mode of an excitonic condensate in analogy to $Ta_2NiSe_5$. However, the phonon-amplitude coupling in $Ta_2Pd_3Te_5$ cannot originate from a structural phase transition, as no detectable structural phase transition accompanies the excitonic-insulator transition in this material (supplementary materials Sec. 3). In the following, we are proposing two possible models to explain the lifetime enhancement below $T_c$ through the coupling exciton-lattice contribution. The first is based on dressing of the coherent phonon by fluctuations of the excitonic order parameter, whereas the second describes the low-temperature lifetime phenomenologically in terms of a renormalized anharmonic phonon background together with a conductivity-dependent electronic damping channel associated with the opening of the excitonic gap.

For $Ta_2NiSe_5$ a picture of polaronic couplings of the EI state to the lattice is proposed based on experimental fingerprints [27, 30, 35, 52]. We argue that also in $Ta_2Pd_3Te_5$ such exciton–polaron complexes, arising from local lattice distortions, may give rise to the coupled phonon–amplitude modes. Within this scenario, the complexes may account for the observed temperature dependence of the coherent amplitude response and could contribute to the exceptionally long mode lifetime by shielding them from the environment [53]. In other words, within this picture the local lattice distortion of the polaron complex may act like a “natural cavity” with the corresponding lifetime enhancement. To quantify this, we introduce a toy model with the concept of an "order-parameter polaron" (see supplementary materials Sec. 5). In this framework, the primary oscillator couples nonlinearly to amplitude fluctuations of an exciton condensate, a mechanism analogous to a standard polaron where a particle is dressed by lattice distortions. By transforming this coupled system into an independent-boson representation, the model demonstrates that the coupling to the condensate fluctuations significantly suppresses the oscillator's interaction with the surrounding acoustic phonon bath. At low temperatures, this dressing effectively shields the coherent mode, leading to an exponential increase in its lifetime. Taking the dressed-mode frequencies to be $\Omega_{pol}$ = 0.61 THz and 0.85 THz, the order-parameter polaron model quantitatively reproduces the observed lifetime enhancement for both coherent modes (solid blue line in Fig. 3(a) for the 0.61 THz mode; see supplementary materials Sec 5 for the 0.85 THz mode). The resulting values $\Omega_b \approx 32.3$ meV for the 0.61 THz mode and $\Omega_b \approx 35.2$ meV for the 0.85 THz mode describe high-frequency fluctuations close to the excitonic gap ($\Delta \approx 40$ meV) [39]. While these values allow understanding the possibility of lifetime enhancement due to hybridization, we note that the experimentally small phonon anharmonicity (Fig. 3(b)) speaks against a strong coupling.

In the second approach, we show that a simple phenomenological model of coupled phonon-exciton (Ph-X) character can describe the phonon-amplitude lifetime enhancement. The anharmonic phonon decay with a reduced electronic scattering rate in the excitonic insulator phase (supplementary materials Sec 5d) can be described as follows:

$$\Gamma(T) = \Gamma_0 + C\left(1 + \frac{2}{e^{\frac{\hbar\omega_0}{2k_BT}} - 1}\right) + \frac{B}{\rho(T)}$$

where $\Gamma_0$ represents the intrinsic phonon scattering rate at zero temperature, and C is the anharmonicity coefficient and B is a phenomenological coefficient that determines the magnitude of the conductivity-dependent electronic contribution to the phonon damping and ρ(T) is the measured electrical resistivity. Below $T_C$, the opening of the excitonic gap increases the resistivity (ρ(T)) and reduces the conductivity, thereby suppressing the electronic contribution to the phonon damping. The resulting model reproduces the temperature dependence of the phonon lifetime below $T_C$, as shown by the solid green line in Fig. 3(a).

**Conclusions:**

We have successfully shown that we can characterize the EI state in $Ta_2Pd_3Te_5$ via the coherent response of coupled phonon-amplitude modes. We find that exciton condensation sets in around the transition temperature of 100 K in agreement with the gap opening found in tunnelling experiments [39] and far below the metal-insulator transition at 365 K [37]. As a novel signature of the condensate phase, we observe nearly undamped coherent oscillations that set in below $T_C$. To understand their remarkable lifetime enhancement, we discuss the dressing of the coherent oscillations by condensate fluctuations within exciton–polaron complexes or phenomenologically by a renormalized anharmonic phonon background together with a conductivity-dependent electronic damping channel associated with the excitonic gap opening.

The stabilization of an excitonic condensate in real materials is often thought to require coupling between excitons and lattice degrees of freedom. In the prototypical excitonic insulator $Ta_2NiSe_5$ this may be realized by the coinciding EI and structural phase transition [34-36, 52]. In contrast, in $Ta_2Pd_3Te_5$ no structural transition is detectable. Our observations suggest that substantial exciton–lattice coupling and exceptionally long-lived coherent excitations emerge even in the absence of a macroscopic structural instability. Whether dominant exciton fields induce the lattice distortion or whether the complexes form based on pre-existing local distortions remains to be tested in future experiments. In this context $Ta_2Pd_3Te_5$ serves as an ideal model system to explore the coherent phonon-amplitude responses under external control parameters.

**Supplementary information**

The Supplementary Material includes experimental details (Supplementary Section 1); additional information on sample growth and characterization (Supplementary Section 2); details of the pump–probe data analysis and fitting procedures (Supplementary Section 3); a discussion of the temperature-dependent behavior of the 0.85 THz mode (Supplementary Section 4); models describing the temperature-dependent coherent-mode dynamics (Supplementary Section 5); and a discussion of the fluence dependence of the 0.61 THz mode (Supplementary Section 6).

**Acknowledgments**

We acknowledge support from the Deutsche Forschungsgemeinschaft (DFG) through the Wurzburg-Dresden Cluster of Excellence on Complexity, Topology and Dynamics in Quantum Matter—ctd. qmat (EXC 2147, Project No. 390858490), Deutsche Forschungsgemeinschaft (DFG), SFB 1143 (project id 247310070), and the DRESDEN-concept alliance of research institutions. A.K.N.M. acknowledges financial support from the Alexander von Humboldt Foundation through the Humboldt Research Fellowship Programme.  S.K. acknowledges funding by the European Union (ERC, T-Higgs, GA 101044657). (Views and opinions expressed are however, those of the author(s) only and do not necessarily reflect those of the European Union or the European Research Council Executive Agency. Neither the European Union nor the granting authority can be held responsible for them). S.C., C.S., and C.F. acknowledge support from Deutsche Forschungsgemeinschaft (DFG) via QUAST (Project ID FOR 5249), and EXQIRAL (No. 101131579). A.R. acknowledge support from Deutsche Forschungsgemeinschaft (DFG) via CRC 1238 (Project No. 277146847, subprojects C02 and C04). M.J.K. acknowledges support from the Brain Pool Program through the National Research Foundation of Korea (NRF), funded by the Ministry of Science and ICT (RS-2024-00407957).

## Supplemental Material for:

# Extraordinary Lifetime Enhancement of Coherent Phonon-Amplitude Modes in the Excitonic Insulator Phase of $Ta_2Pd_3Te_5$

**Anjan Kumar N M[1*], Shuhan Wang[1], Snehashish Chatterjee[2], Yan Zhu[1], MinJae Kim[1,3], Tobias Ritschel[1], Elaheh Sadrollahi[1], Jochen Geck[1,5], Achim Rosch[4], Chandra Shekhar[2], Claudia Felser[2,5], Stefan Kaiser[1,5*]**

[1] Institute of Solid State and Materials Physics, Technische Universität, Dresden, Germany.

[2] Topological Quantum Chemistry Group, Max Planck Institute for Chemical Physics of Solids, Dresden, Germany

[3] Korea Institute of Science and Technology, Seoul, South Korea.

[4] Institute for Theoretical Physics, University of Cologne, Cologne, Germany.

[5] Würzburg-Dresden Cluster of Excellence ctd.qmat, Technische Universität Dresden, 01062 Dresden, Germany

*Email: anjan.naralapura@tu-dresden.de ; stefan.kaiser@tu-dresden.de

**Contents:**

## 1. Time-resolved reflectivity measurements

Time-resolved experiments use 190 fs laser pulses in reflection geometry with 1030 nm pump and 680 nm probe wavelength. They are generated from regenerative amplifier system (Pharos, Light Conversion; 1030 nm central wavelength, 10 kHz repetition rate, 2 mJ pulse energy) and a corresponding optical parametric amplifier. Pump and probe beams are linearly polarized with mutually perpendicular polarizations to minimize coherent artifacts and polarization-dependent interference effects. The relative time delay between the pump and probe pulses is controlled using a motorized delay stage providing up to 600 ps of temporal delay. The pump beam is modulated using a mechanical chopper, and the pump-induced change in reflectivity, $\Delta R/R$, is detected using a photodiode via lock-in detection. Temperature-dependent measurements use a cold-finger helium cryostat (cryoVac) with the sample in vacuum, ensuring an inert environment and pristine sample surface throughout the experiment. Spot sizes of the pump and probe beams are 220 $\mu m$ and 120 $\mu m$ (FWHM), respectively, ensuring homogeneous excitation within the probed region. For all the measurements the probe fluence is kept at 5 µJ/cm$^2$. For the pump-fluence-dependent measurements, the pump fluence is varied from 0.03 to 1.27 mJ/cm$^2$ while maintaining the same experimental configuration.

## 2. Crystal growth and characterization

High-quality single crystals of $Ta_2Pd_3Te_5$ were synthesized using a chemical vapor transport technique. Stoichiometric amounts of high-purity elemental Ta, Pd, and Te were thoroughly mixed, followed by the addition of iodine as a transport agent (~5 mg/cm$^3$). The mixture was sealed in an evacuated quartz ampoule under an inert argon atmosphere and subjected to a controlled temperature gradient, with the hot and cold zone maintained at 850 °C and 800 °C respectively, for crystal growth over an extended period of 2 weeks. Needle-like exfoliable single crystals of few mm in length were obtained at the cooler end of the quartz ampoule. For sample characterization, energy-dispersive X-ray spectroscopy (EDS) measurements were carried out. Figure S1 presents the EDX pattern, from which the elemental composition was determined to be Ta: Pd: Te = 2: 2.64: 4.62.

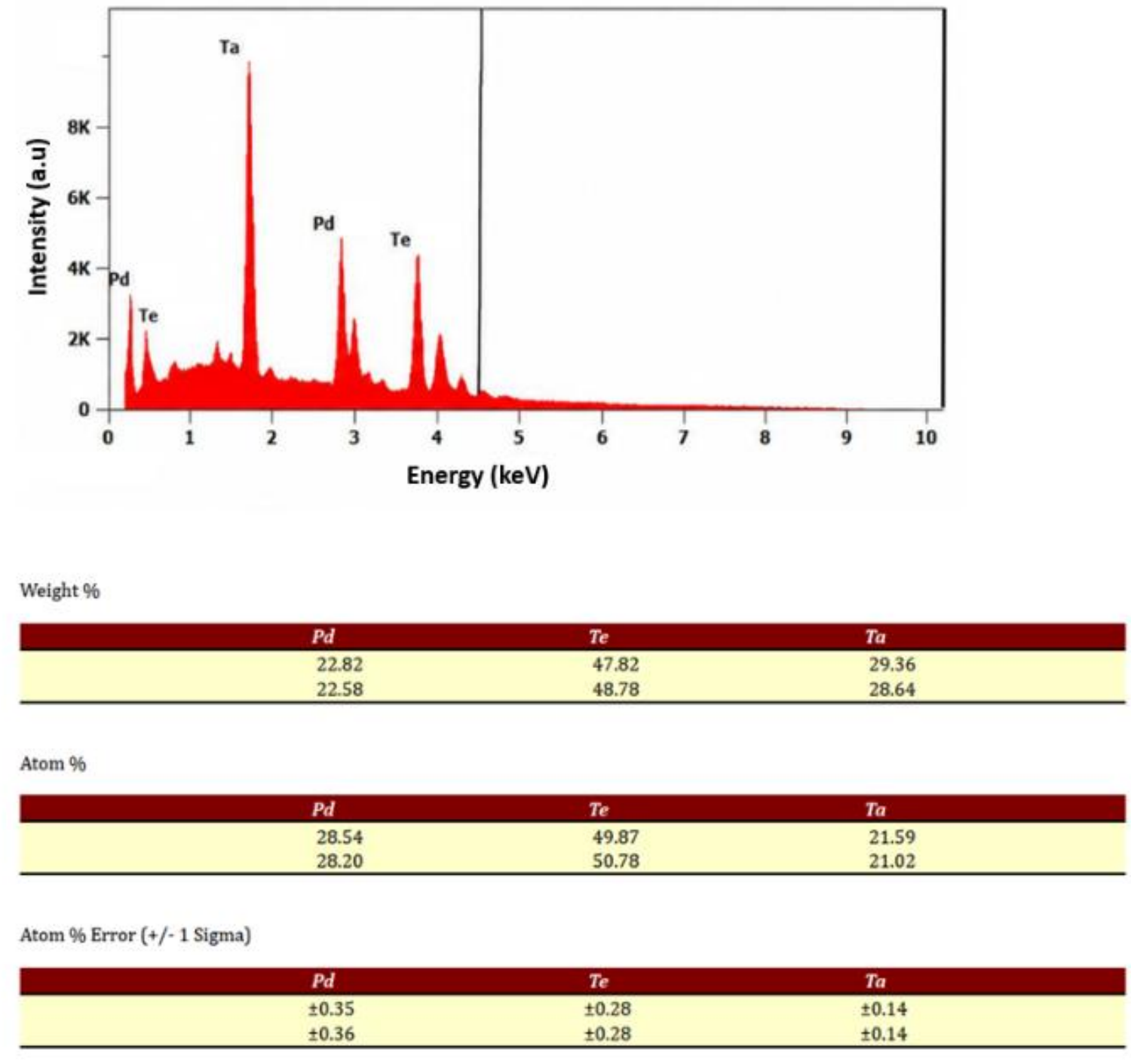

Weight %

| Pd | Te | Ta |
|---|---|---|
| 22.82 | 47.82 | 29.36 |
| 22.58 | 48.78 | 28.64 |

Atom %

| Pd | Te | Ta |
|---|---|---|
| 28.54 | 49.87 | 21.59 |
| 28.20 | 50.78 | 21.02 |

Atom % Error (+/- 1 Sigma)

| Pd | Te | Ta |
|---|---|---|
| ±0.35 | ±0.28 | ±0.14 |
| ±0.36 | ±0.28 | ±0.14 |

Figure S1: **EDX Characterization of $Ta_2Pd_3Te_5$.** EDX pattern and the elemental composition of $Ta_2Pd_3Te_5$ crystal.

**Transport data:**

Temperature-dependent zero-field resistivity measurements were performed using a standard four-probe method in a Physical Property Measurement System (PPMS). As shown in Fig. S2a, the R–T data display semiconducting behaviour. In semiconductors, charge transport is typically governed by thermal activation and can be described by $\rho = \rho_0 \exp\left(-\frac{E_A}{2k_B T}\right)$, where $\rho_0$ is a prefactor, $E_A$ represents the activation energy, and $k_B$ is the Boltzmann constant. The activation energy was determined by fitting the resistivity data using the Arrhenius relation, $E_A = -2k_B T^2 \left(-\frac{d(\ln\rho)}{dT}\right)$, and its temperature dependence is presented in Fig. S2 (b). The extracted values are in good agreement with values reported in literature [1].

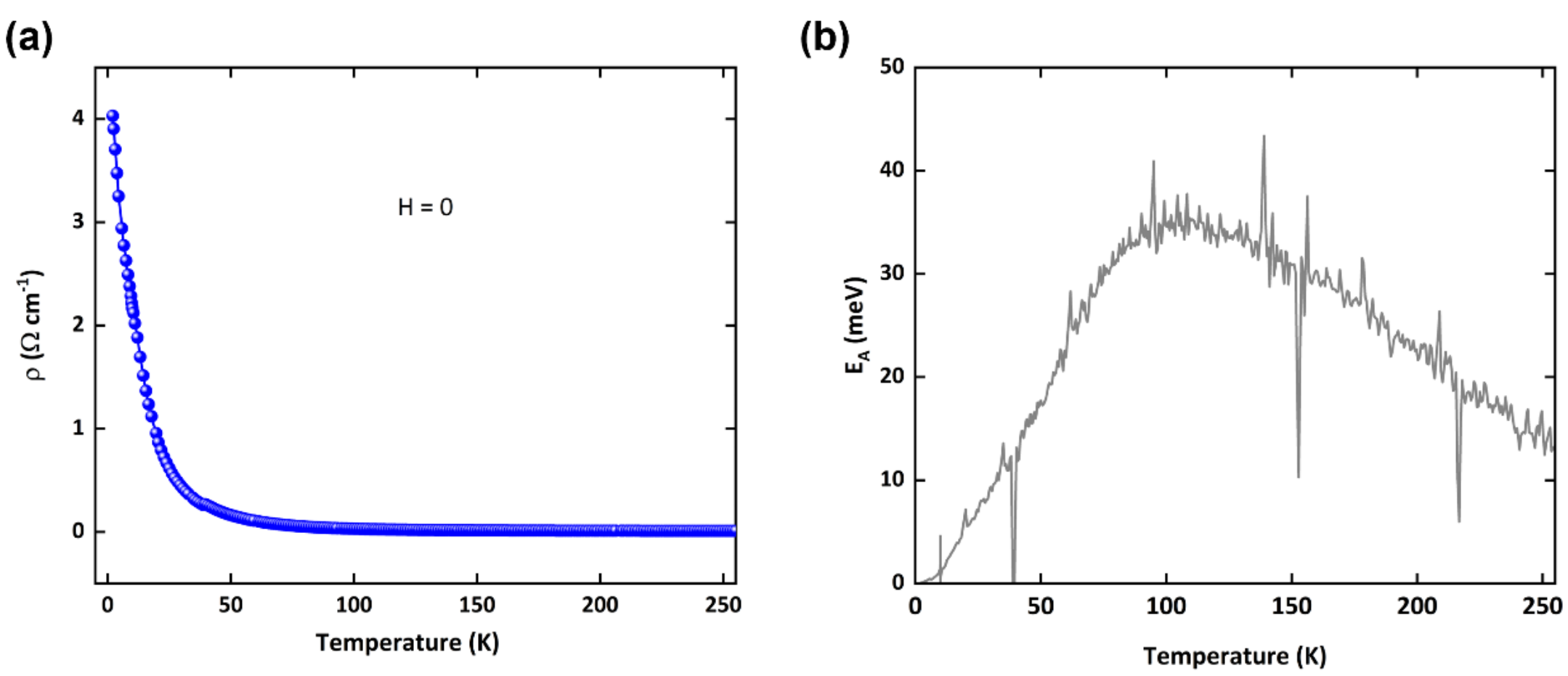

Figure S2: **Temperature-Dependent Resistivity and Activation Energy of $Ta_2Pd_3Te_5$**. Resistivity of $Ta_2Pd_3Te_5$ as a function of temperature and its corresponding activation energy (b) extracted from Arrhenius equation.

**Single- crystal X- ray diffraction at 10 K and 290 K: Measurement and refinement**

Single-crystal X-ray diffraction data were collected on the same $Ta_2Pd_3Te_5$ crystal at 290 K and at 10 K on a custom-built in-house six-circle diffractometer. X-rays were produced by an Excillum liquid-metal-jet source using an In–Ga alloy anode, from which the In $K\alpha_{1,2}$ radiation (λ = 0.51341 Å) was selected and focused by a multilayer mirror. The sample was mounted in a closed-cycle pulse-tube cryostat and the diffracted intensities were recorded with a Dectris Pilatus 1M hybrid pixel-array detector. At each temperature three ω-scan runs (0.5° steps) were recorded. The frames were indexed, integrated and corrected for absorption with a multi-scan (spherical-harmonics, SCALE3 ABSPACK) procedure in CrysAlisPRO. The structures were solved with SHELXT and refined against $F^2$ by full-matrix least squares with SHELXL within the Olex2 package.

At both temperatures we found an orthorhombic space group Pnma (No. 62) with Z = 4 and an asymmetric unit comprising two Ta, three Pd and five Te sites, all on the mirror plane at y = 1/4 or 3/4 (Fig. S4). In the structure the Ta atoms condense into ladders running along the short b axis (Fig. S4 (b)). The refinements converged to low residuals ($R_1$ = 0.0136 at 10 K and 0.0185 at 290 K) using 33 parameters and no restraints. Comparison of the two data sets shows that the structure at 10 K and 290 K is essentially identical: the atomic fractional coordinates agree within their standard uncertainties (Table S2), the Ta–Te, Pd–Te and Pd–Pd bond lengths and the coordination geometry are unchanged, and the space group is retained. The only significant difference is the expected thermal contraction of the lattice on cooling all three cell edges shorten slightly and the cell volume decreases by about 0.8% (from 958.4(10) to 951.0(6) $Å^3$), accompanied by the corresponding reduction of the atomic displacement parameters. No structural phase transition is observed between 290 K and 10 K. The most important crystallographic and refinement quantities for both temperatures are summarised in Table S1, and the quality of the fits is illustrated by the observed-versus-calculated structure-factor plots in Fig. S3.

Table S1: **Single-crystal X-ray diffraction refinement parameters**. Crystallographic data and refinement details for $Ta_2Pd_3Te_5$ at 10 K and 290 K

| Parameter | 10 K | 290 K |
|---|---|---|
| Empirical formula | $Pd_3Ta_2Te_5$ | $Pd_3Ta_2Te_5$ |
| Formula weight (g mol$^{-1}$) | 1319.10 | 1319.10 |
| Crystal system | orthorhombic | orthorhombic |
| Space group | Pnma (No. 62) | Pnma (No. 62) |
| Z | 4 | 4 |
| Radiation, λ (Å) | In $K\alpha_{1,2}$, 0.51341 | In $K\alpha_{1,2}$, 0.51341 |
| a (Å) | 13.881(9) | 13.902(15) |
| b (Å) | 3.6970(2) | 3.7061(4) |
| c (Å) | 18.5303(14) | 18.602(2) |
| V (Å$^3$) | 951.0(6) | 958.4(10) |
| $\rho_{calc}$ (g cm$^{-3}$) | 9.214 | 9.142 |
| μ (mm$^{-1}$) | 18.261 | 18.120 |
| θ range (°) | 2.61–23.89 | 2.60–23.90 |
| Reflections collected | 621 | 636 |
| Independent reflections | 308 | 317 |
| $R_{int}$ | 0.0131 | 0.0096 |
| Reflections $I > 2\sigma(I)$ | 275 | 282 |
| Parameters / restraints | 33 / 0 | 33 / 0 |
| $R_1$ [$I > 2\sigma(I)$] | 0.0136 | 0.0185 |
| $wR_2$ (all data) | 0.0339 | 0.0475 |
| Goodness-of-fit on $F^2$ | 1.101 | 1.046 |
| $\Delta\rho_{max}/\Delta\rho_{min}$ (eÅ$^{-3}$) | 0.66 / −0.61 | 0.78 / −0.84 |

Table S 2 : **Atomic fractional coordinates of $Ta_2Pd_3Te_5$:** Wyckoff positions and fractional atomic coordinates of $Ta_2Pd_3Te_5$ at 10 K and 290 K. In space group Pnma (No. 62) all ten atoms occupy the special Wyckoff position 4c (site symmetry.m.) on the mirror plane at y = 1/4 or 3/4. Standard uncertainties are given in parentheses.

| Atom | Wyckoff | 10 K | | | 290 K | | |
|---|---|---|---|---|---|---|---|
| | | x | y | z | x | y | z |
| Ta1 | 4c | 0.74936(9) | 0.7500 | 0.91684(3) | 0.74898(12) | 0.7500 | 0.91702(3) |
| Ta2 | 4c | 0.75658(9) | 0.2500 | 0.66696(3) | 0.75653(12) | 0.2500 | 0.66690(3) |
| Te3 | 4c | 0.59938(15) | 0.7500 | 0.44092(4) | 0.59950(19) | 0.7500 | 0.44101(6) |
| Te4 | 4c | 0.61296(17) | 0.2500 | 0.86088(4) | 0.6133(2) | 0.2500 | 0.86104(5) |
| Te5 | 4c | 0.86763(16) | 0.2500 | 0.54049(4) | 0.8677(2) | 0.2500 | 0.54043(5) |
| Te6 | 4c | 0.60478(18) | 0.7500 | 0.64638(4) | 0.6051(2) | 0.7500 | 0.64634(6) |
| Te7 | 4c | 0.60547(15) | 0.2500 | 0.22009(4) | 0.6057(2) | 0.2500 | 0.22021(6) |
| Pd1 | 4c | 0.731(2) | 0.7500 | 0.54193(5) | 0.7313(3) | 0.7500 | 0.54201(7) |
| Pd2 | 4c | 0.6915(2) | 0.7500 | 0.77258(5) | 0.6918(3) | 0.7500 | 0.77256(7) |
| Pd3 | 4c | 0.6783(2) | 0.7500 | 0.31134(5) | 0.6785(3) | 0.7500 | 0.31156(7) |

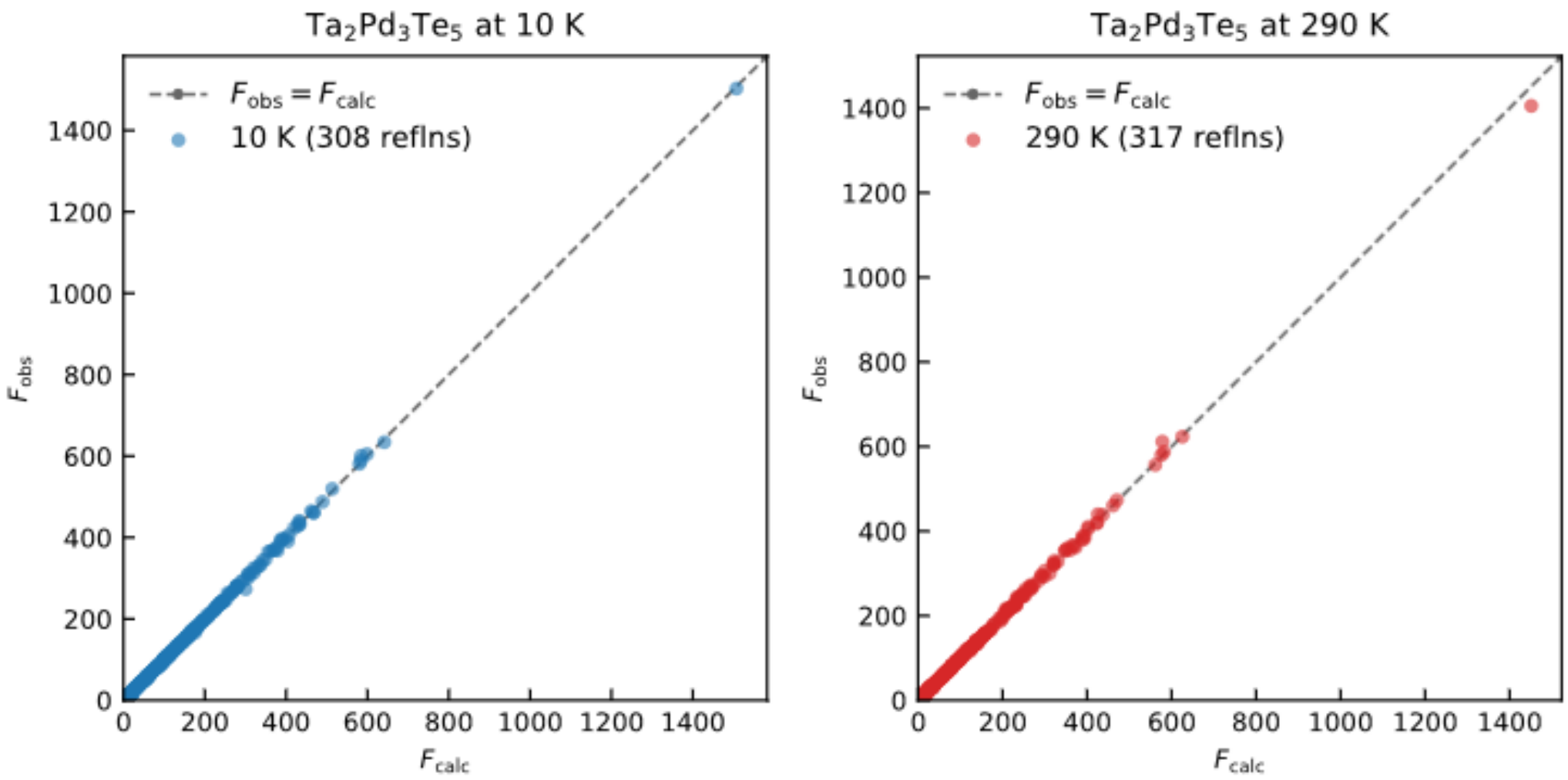

Figure S3: **Observed versus calculated structure-factor amplitudes ($F_{obs}$ vs. $F_{calc}$) for the $Ta_2Pd_3Te_5$ refinements at 10 K (left, 308 reflections) and 290 K (right, 317 reflections)**. Amplitudes were obtained as $F = \sqrt{F^2}$ from the SHELXL. fcf output. The dashed line marks the ideal relation $F_{obs} = F_{calc}$; at both temperatures the data scatter closely around it, consistent with the low residuals of the refinements ($R_1$ = 0.0136 and 0.0185, respectively).

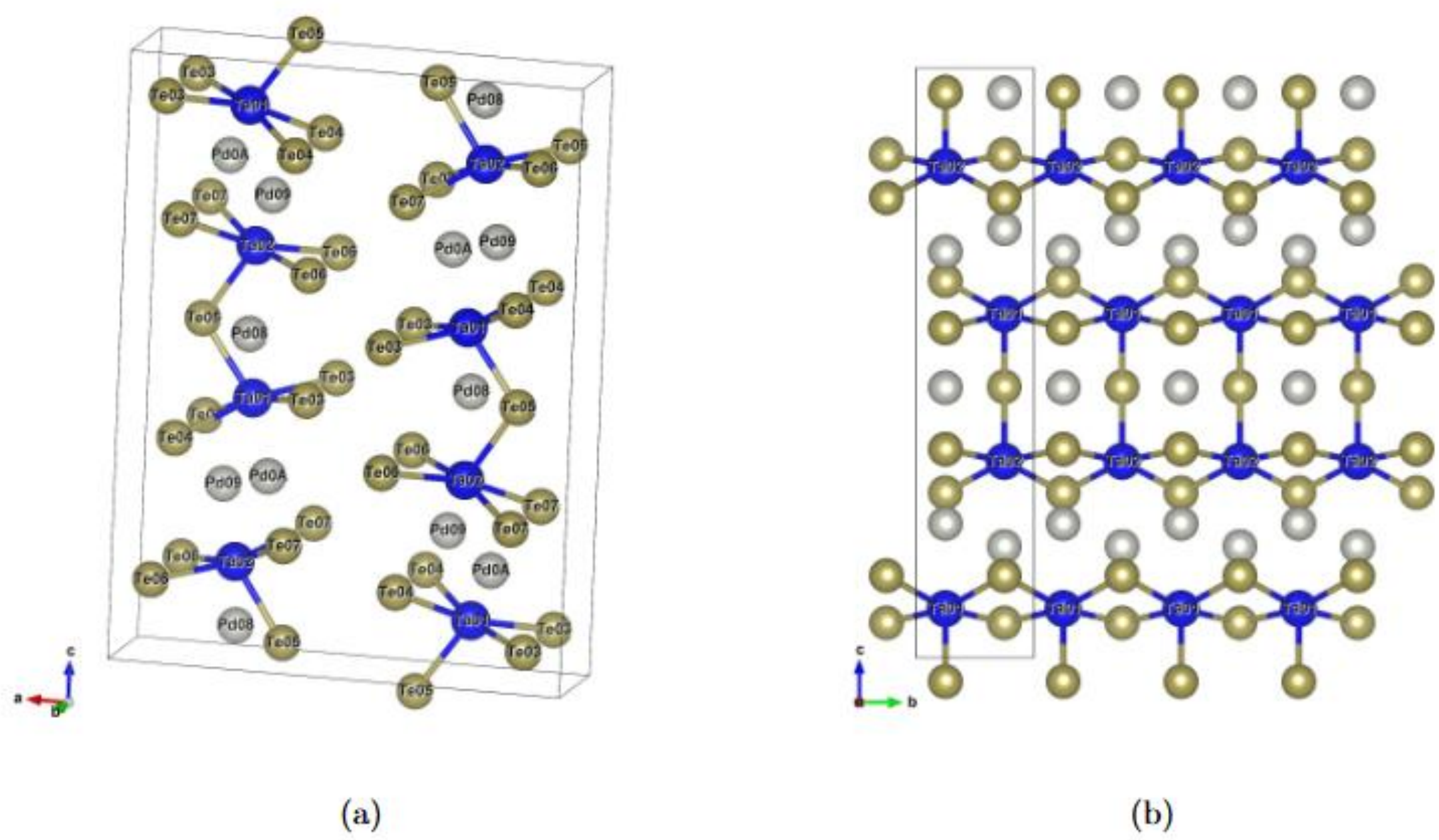

Figure S 4: **Crystal structure of $Ta_2Pd_3Te_5$ (space group Pnma).** (a) View of a single unit cell with the symmetry-independent atoms labelled (Ta, Pd and Te). (b) Extended view spanning 0.5 a × 4 b × 1 c, in which the Ta atoms condense into the characteristic Ta ladders running along the short b axis.

## 3. Data analysis:

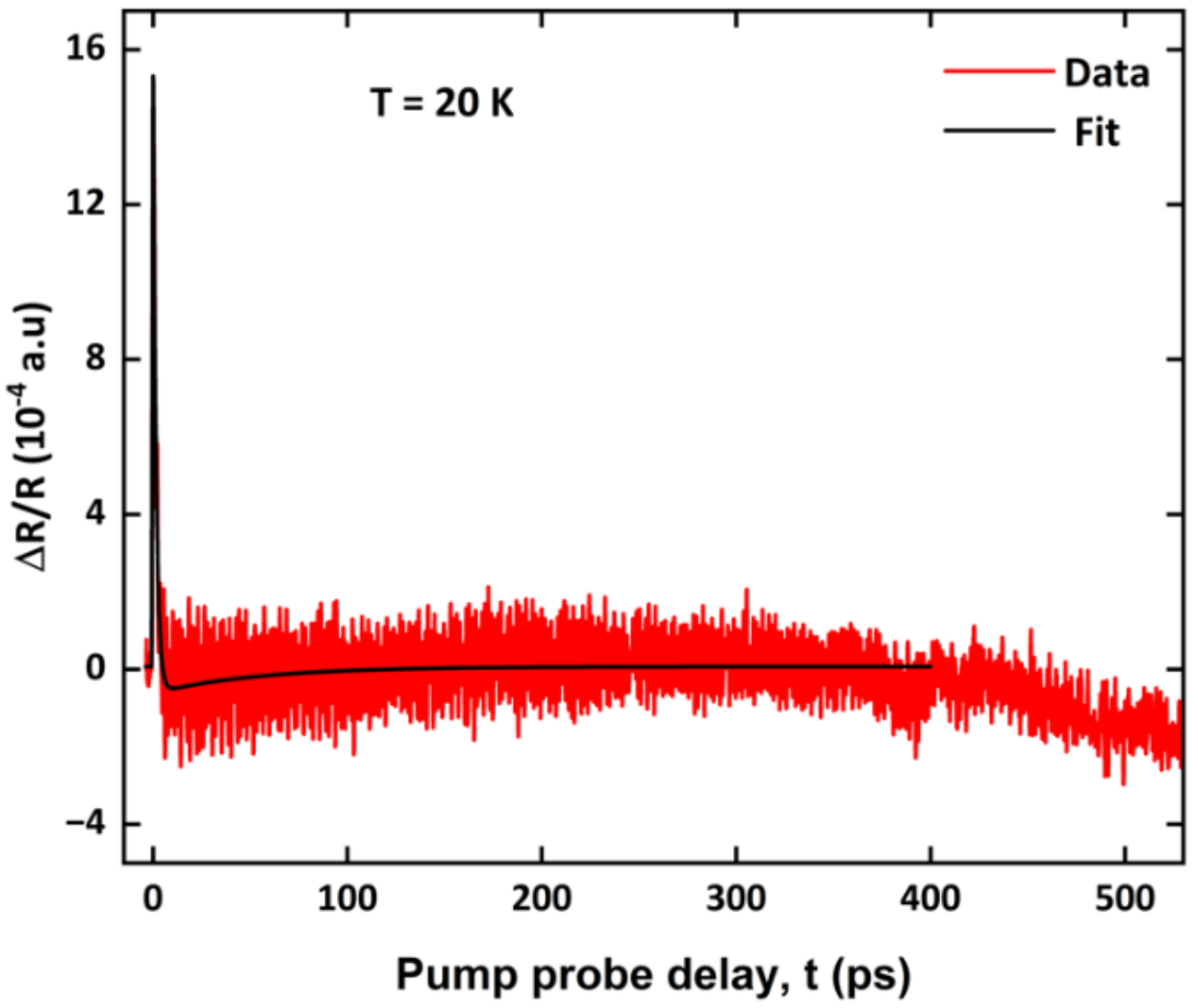

Figure S5: **Photoinduced Reflectivity Dynamics of $Ta_2Pd_3Te_5$.** Photoinduced reflectivity change ($\Delta R/R$) at 20 K as a function of pump–probe delay time t, shown up to 530 ps. The data is fitted (black solid line) up to 400 ps to equation S1. The measurements were performed using a pump fluence of 20 µJ/cm$^2$.

Photoinduced reflectivity changes are recorded over a time window of 530 ps, as shown in Fig. S5. The signal shows an electronic background with superimposed long-lived coherent oscillations persisting up to 530 ps. Beyond 400 ps the signal becomes slightly affected by alignment instabilities of the mechanical stage for long delay times leading to a slight loss of perfect spatial pump-probe overlap.

That limits our available experimental time window to measure the damping of the coherent oscillations to about 400 ps to be used in the fit of the oscillations. Nevertheless, that window is sufficient to trace the remarkable lifetime enhancement of the oscillations of 654 ps at 20 K in $Ta_2Pd_3Te_5$, far beyond the longest coherent phonon lifetime reported in literature of 211 ps at 5 K in ZnO [2].

**Electronic background:**

The electronic background of the ΔR/R signal exhibits a prompt rise immediately after photoexcitation, followed by a bi-exponential decay that we can describe with the typical form of [3, 4]

$$\frac{\Delta R(t)}{R} = \left[A_1 e^{-\frac{(t-t_C)}{\tau_1}} + A_2 e^{-\frac{(t-t_C)}{\tau_2}} + B\right] \times \left[\mathrm{erf}\left(\frac{(t-t_C)}{t_r}\right) + 1\right] \qquad Eq.\ S1$$

where the error function describes the sudden onset within a rise time $t_r$ of 0.27 ± 0.02 ps that remains nearly independent of temperature and excitation fluence. $A_1$ and $A_2$ are the amplitudes of the fast carrier-carrier and slow electron–phonon relaxation channels, respectively, with corresponding relaxation times $\tau_1$ and $\tau_2$, while B denotes a constant background. $\boldsymbol{t_c}$ denotes the temporal overlap (time zero) between pump and probe pulses.

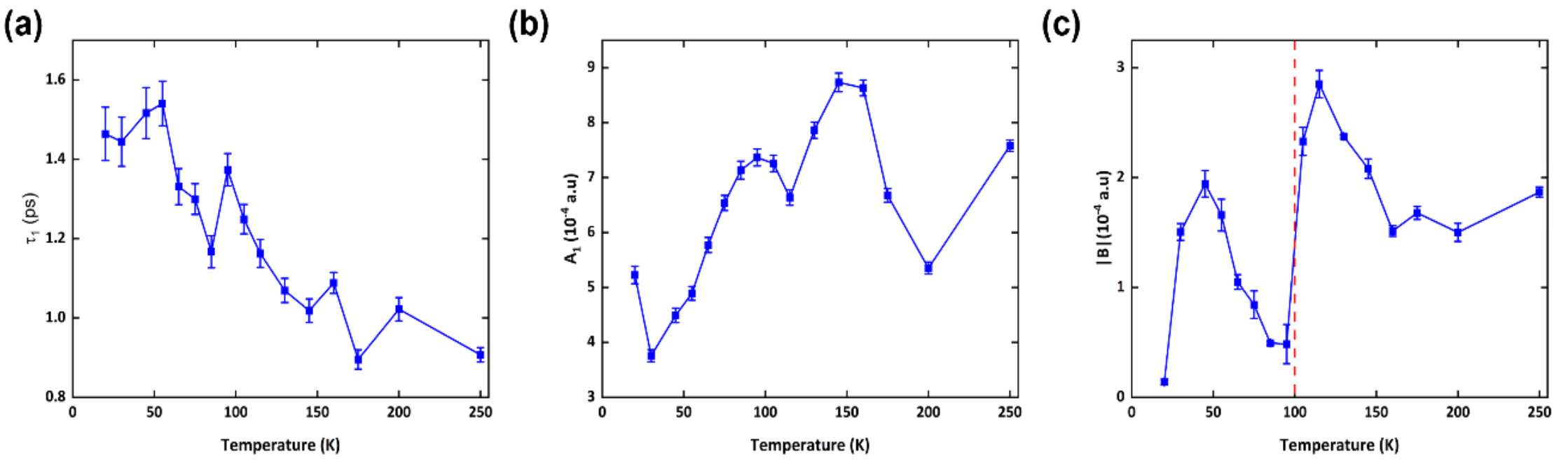

Figure S6: **Temperature dependence of the electronic relaxation parameters at an excitation density of 20 μJ/cm².** (a) Fast relaxation time $\tau_1$ and (b) its corresponding amplitude $A_1$. (c) Amplitude, B of the long-lived component. The dashed red line indicates the $T_c$.

Their temperature dependencies are shown in Fig. S6. The fast relaxation time, $\tau_1$ (Fig. S6 a), is approximately 1.5 ps at low temperature and decreases monotonically with increasing temperature, showing a small anomaly around the transition temperature of 100 K. Previous studies have suggested that the pump pulse transiently dissociates bound excitons, generating a non-equilibrium carrier distribution. Subsequently, $\tau_1$ captures carrier cooling and recombination into excitons [3, 4], which are mainly governed by carrier–carrier interactions. The corresponding amplitude $A_1$ (Fig. S6b) shows a non-monotonic behaviour around 100 K. This may point to a crossover regime around the transition temperature. Exploring this is beyond the scope of the present manuscript. The slow relaxation time $\tau_2$ and its corresponding amplitude $A_2$, associated with carrier–phonon interactions such as exciton–phonon coupling, capture the phase transition with sharp features as discussed in the main manuscript (Fig. 1c). The same is found for the long-lived component B shown in Fig. S6c showing a sharp drop at $T_C$ = 100 K.

**Oscillatory signal:**

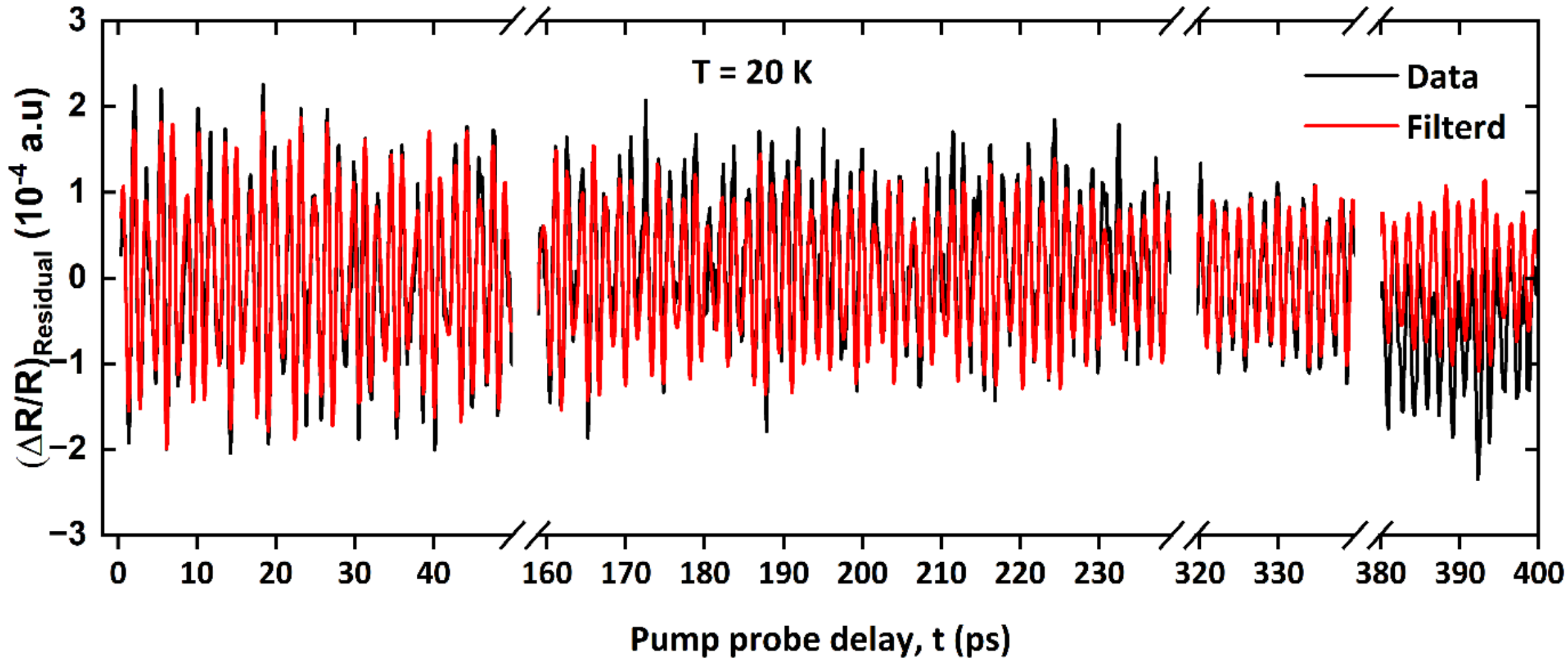

Figure S7 : **Coherent mode oscillations in $Ta_2Pd_3Te_5$.** Residual oscillatory components (black solid line) at 20 K obtained after subtracting the electronic background from the transient reflectivity ($\Delta R/R$) and the filtered data using the bandpass filter (red solid line) shown up to 400 ps.

Subtracting the electronic background (Eq. S1) from the measured signal reveals the raw oscillations as shown in Fig. S7 (black).

The corresponding FFT spectrum in Fig. S8 shows three phonon modes at 0.61, 0.85 and 1.03 THz. The 1 THz mode is just above the noise level. Therefore we focus on the other two modes and can apply a band-pass filter of 0.55–0.9 THz to the data (Figs. S7 and S8, red): The filter does not affect the oscillatory features or frequency components of our two modes of interest.

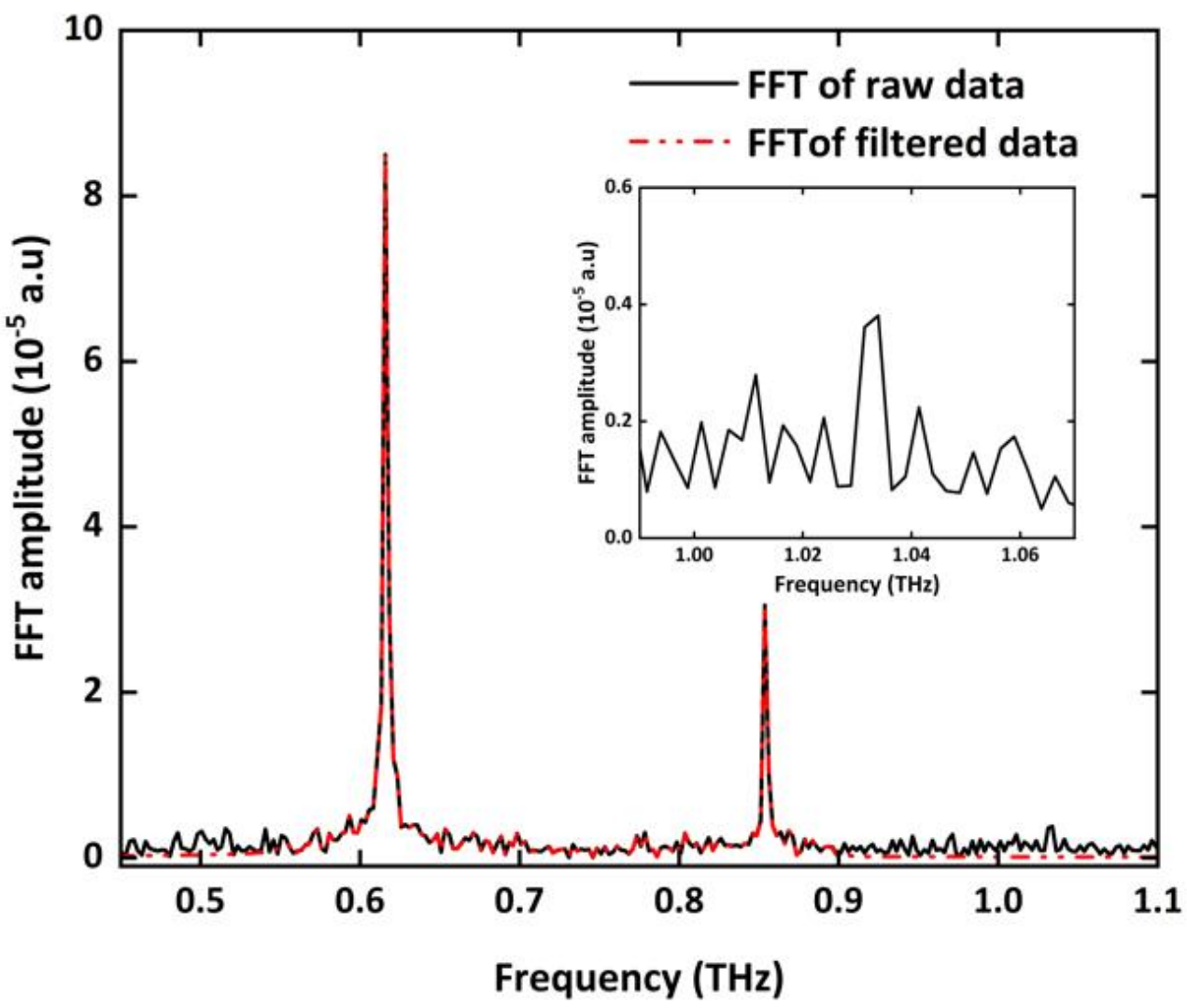

Figure S8: **FFT Spectra of Raw and Filtered Residual Oscillations**. The fast Fourier transforms (FFTs) spectra of the raw and filtered residual time domain oscillations shown in S7.

These filtered time-domain oscillations are fit using the equation

$$\left(\frac{\Delta R(t)}{R}\right)_{Osc} = A_1^{ph} e^{-\left[\frac{t}{\tau_1^{ph}}\right]} \cos\left(2\pi\nu_1^{ph} t + \phi_1^{ph}\right) + A_2^{ph} e^{-\left[\frac{t}{\tau_2^{ph}}\right]} \cos\left(2\pi\nu_2^{ph} t + \phi_2^{ph}\right) \qquad \text{Eq. S2}$$

where $A_i^{ph}$ are the overall amplitudes, $\tau_i^{ph}$ and $\nu_i^{ph}$ are the oscillation lifetimes and frequencies and $\phi_i^{ph}$ are their phases.

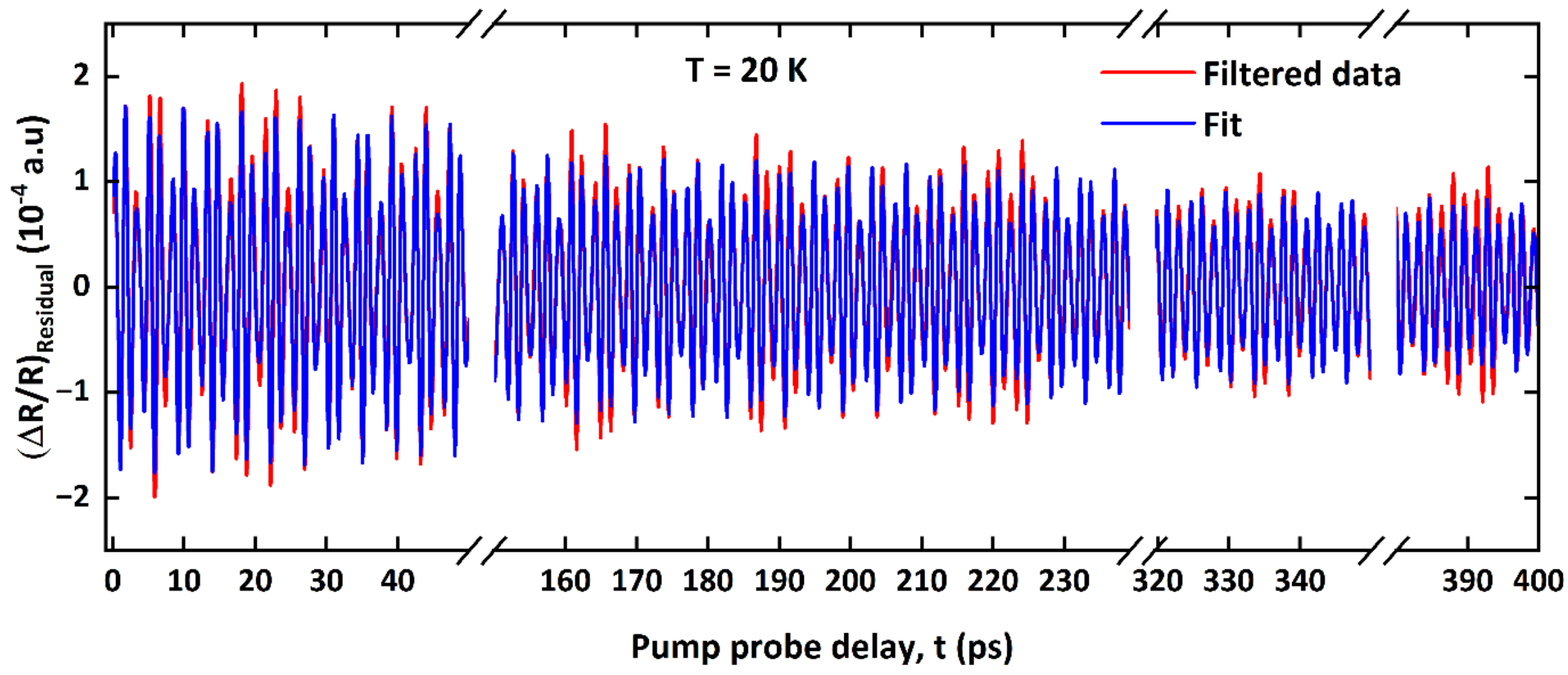

Figure S9: **Fit to the Filtered Residual Oscillations.** The filtered residual oscillations (red solid line) at 20 K and its fit (blue solid line) to Eq. S2.

The filtered oscillations (red) and the corresponding fit (blue) up to 400 ps are shown in Fig. S9. The FFTs from these time traces up to 400ps are shown in Fig. 2 (b) of the manuscript for all temperatures while the time-domain data in the main text Fig. 2(a) is only shown up to 100 ps for a clear visualisation of the oscillations and damping at higher temperatures.

## 4. 0.85 THz mode

The temperature dependence of the amplitude, lifetime, and frequency of the 0.85 THz mode (Fig. S10) shows a similar behaviour as the 0.61 THz mode discussed in the manuscript.

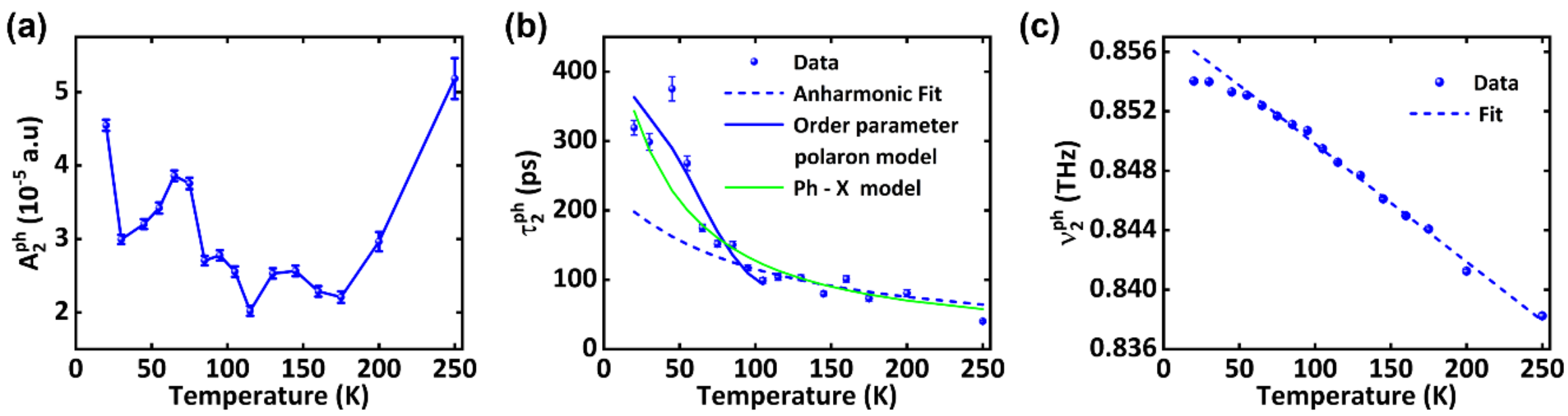

Figure S10: **Temperature-dependent behavior of the 0.85 THz mode at an excitation density of 20 μJ/cm².** (a) Mode amplitude as a function of temperature, showing an enhancement below $T_C$ ~100 K. (c) Mode lifetime vs. temperature, displaying an enhancement below 100 K fitted to an anharmonic model (blue dashes). The behaviour of the model lifetime with temperature is captured using order parameter polaron model (solid blue line) and the coupled phonon-exciton model (Ph-X) (solid green line). (b) Temperature dependent mode frequency. Blue dashes represent the anharmonic fit.

The amplitude of the 0.85 THz mode in Fig. S10(a) increases across $T_C$ and follows a trend similar to that of the gap extracted from ARPES measurements [5]. The same as we found for the 0.61 THz mode discussed in the main text. We note that the amplitudes for both, the 0.61 and 0.85 THz, modes do not show a sharp feature at $T_C$. Instead, a crossover regime between 80 and 120 K of the enhancement is found. As for the crossover regime in the fast component of the electronic signal, and the potential link to the gap evolution in the ARPES probes, the details require further studies that are beyond the scope of the present manuscript. The lifetime of the 0.85 THz mode in Fig. S10(b) shows the same characteristic features as the 0.61 THz mode discussed in the main text: A linear temperature dependence above the transition temperature, $T_C$ (blue dashes), as we expect for a coherent phonon mode and a non-linear enhancement below $T_C$ that is not captured by a simple anharmonic phonon model (blue dashes) but is well captured by our exciton-polaron toy model (blue solid line) and coupled phonon-exciton model (Ph-X) (green solid line).We note that a spike around 50 K appears and we also see a remanence of that in the 0.61 THz mode in the main text. However, the potential origins of that are out of the focus of this manuscript and have to be investigated in further studies. The temperature-dependent frequency shift shown in Fig. S10(c) is fitted using the anharmonic phonon model [Eq. (S3)]. Above $T_C$, the model reproduces the observed frequency shift, whereas below $T_C$, the frequency deviates from the anharmonic behavior.

## 5. Models for Temperature-Dependent Lifetime-Enhancement

### a. Anharmonic model.

The temperature dependence of the frequencies of both modes was analysed using a cubic anharmonic decay model in which an optical phonon decays into a pair of acoustic phonons with equal and opposite momenta, q and -q described by [6]:

$$\omega(T) = \omega_0 - A\left(1 + \frac{2}{e^{\frac{\hbar\omega_0}{2k_BT}} - 1}\right) \qquad Eq.S3$$

where $\omega_0$ represents the intrinsic phonon frequency at zero temperature, and A is the anharmonicity coefficient that quantifies the contribution from cubic anharmonic phonon–phonon interactions.

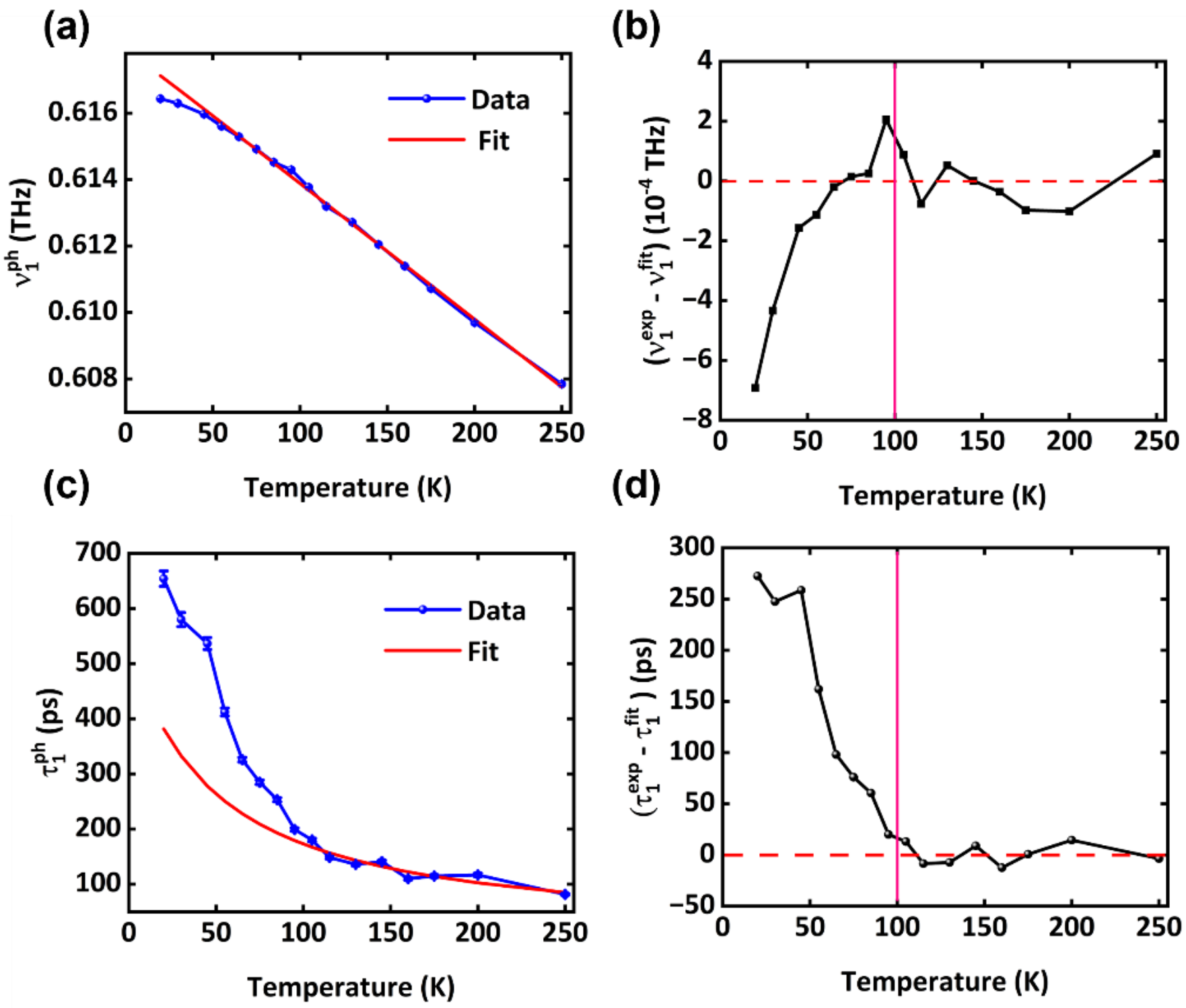

Figure S 11: **Temperature-dependent behavior of the 0.61 THz coherent mode measured at an excitation fluence of 20 μJ/cm².** (a) Temperature dependence of the phonon frequency. The red solid line represents the anharmonic fit. (b) Difference between the measured phonon frequency and the anharmonic fit, highlighting the anomaly across the transition temperature, $T_C$. (c) Temperature dependence of the phonon lifetime. The red solid line is the fit to the anharmonic decay model. (d) Difference between the measured lifetime and the anharmonic fit as a function of temperature, highlighting the onset and evolution of the lifetime enhancement below $T_C$.

The data and the fit of 0.61 THz mode is shown in Fig. S 11 (a) and for the 0.85 THz mode is shown in S10 (c). Above $T_C$, the model captures the temperature evolution of the frequencies with the fitting parameters $\omega_0/2\pi$ = 0.617 THz and A = $3\times10^{-4}$ THz, where as it deviates below $T_C$. The difference in the measured frequency and the fit is shown in Fig S11 (b) for 0.61 THz mode which captures an anomaly across $T_C$ and the deviation from the $T_C$ = 100 K.

The inverse of the lifetime (scattering rate) of both modes was fitted using the same cubic anharmonic decay model, described by:

$$\Gamma(T) = \Gamma_0 + C\left(1 + \frac{2}{e^{\frac{\hbar\omega_0}{2k_BT}} - 1}\right) \qquad Eq.S4$$

where $\Gamma_0$ represents the intrinsic phonon scattering rate at zero temperature, and C is the anharmonicity coefficient.

The experimental data and corresponding fits for the 0.61 THz and 0.85 THz modes are shown in Fig. S11(c) and Fig. S10(b), respectively. The anharmonic model with fit parameters $\Gamma_0$= $1.78\times10^{-3}$ ps$^{-1}$ and C = $2.91\times10^{-4}$ ps$^{-1}$ provides a good description of the temperature dependence above 100 K; however, below the transition temperature, a clear deviation from the model is observed, as shown in Fig. S11(d).

A Raman spectroscopy study [6] on $Ta_2Pd_3Te_5$ reported that the temperature dependence of the scattering rate of the same ($A_g$) phonon mode is well described by an anharmonic model incorporating both cubic and quartic phonon decay processes, with no evidence for anomalous behavior across the transition temperature. Using the reported anharmonic coefficients (C = 0.006) and (D = -3.5 x$10^{-5}$), we simulated the corresponding phonon lifetime. The simulated lifetime together with the coherent phonon lifetime measured in the present work, is presented in Fig. S12. In contrast to the anharmonic prediction, the coherent phonon lifetime measured in this work exhibits a pronounced enhancement below the transition temperature, demonstrating a clear deviation from conventional anharmonic behavior. A similar distinction has been reported for the prototypical excitonic insulator $Ta_2NiSe_5$. Under strong impulsive optical excitation, the 1 THz ($A_g$) phonon exhibits pronounced coupling to the excitonic condensate and acquires amplitude-mode character [7,8]. In contrast, equilibrium Raman measurements show essentially no anomalous change in either the phonon linewidth or frequency across the transition temperature. This comparison highlights that coherent pump–probe spectroscopy can reveal nonequilibrium collective dynamics that are not directly accessible through equilibrium Raman scattering.

### b. **Modified Arrhenius model**:

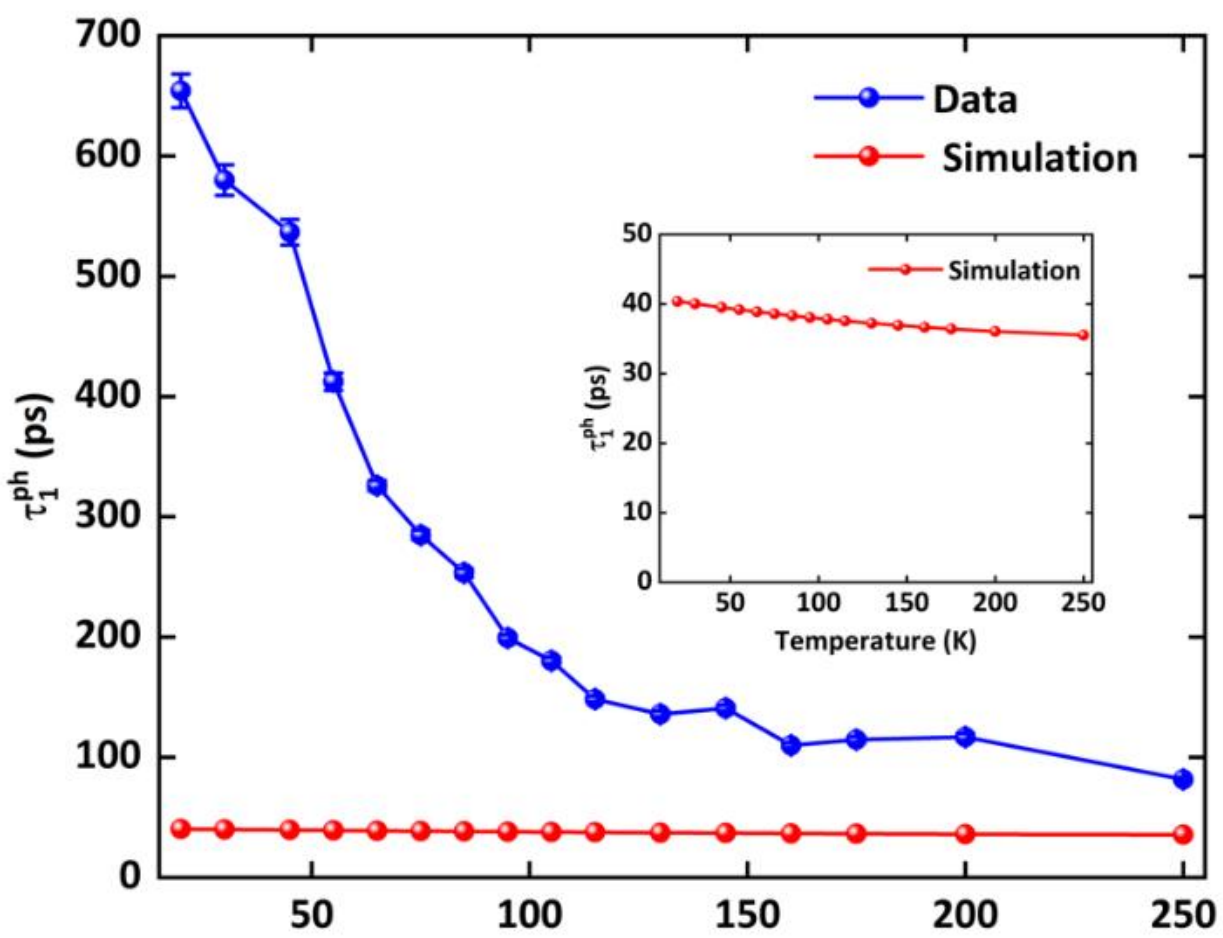

Figure S 12: **Comparison of Experimental and Anharmonic Phonon Lifetimes**. The blue solid line represents the experimentally measured lifetime using pump probe spectroscopy, while the red solid line shows the calculated lifetime of the same ($A_g$) phonon mode based on the anharmonic decay model reported in Ref. *[6]*. The inset shows the latter in a smaller lifetime scale to visualize the slight enhancement on decreasing temperature.

#### b1.

To examine the role of thermally activated electronic damping, we modelled the inverse phonon lifetime using a modified Arrhenius equation,

$$\frac{1}{\tau(T)} = \frac{1}{\tau_0} + A \exp\left(-\frac{E_a}{k_B T}\right) \qquad Eq.S5$$

where $E_a$ is the activation energy, A is a prefactor, and $\frac{1}{\tau_0}$ represents the residual temperature-independent decay rate.

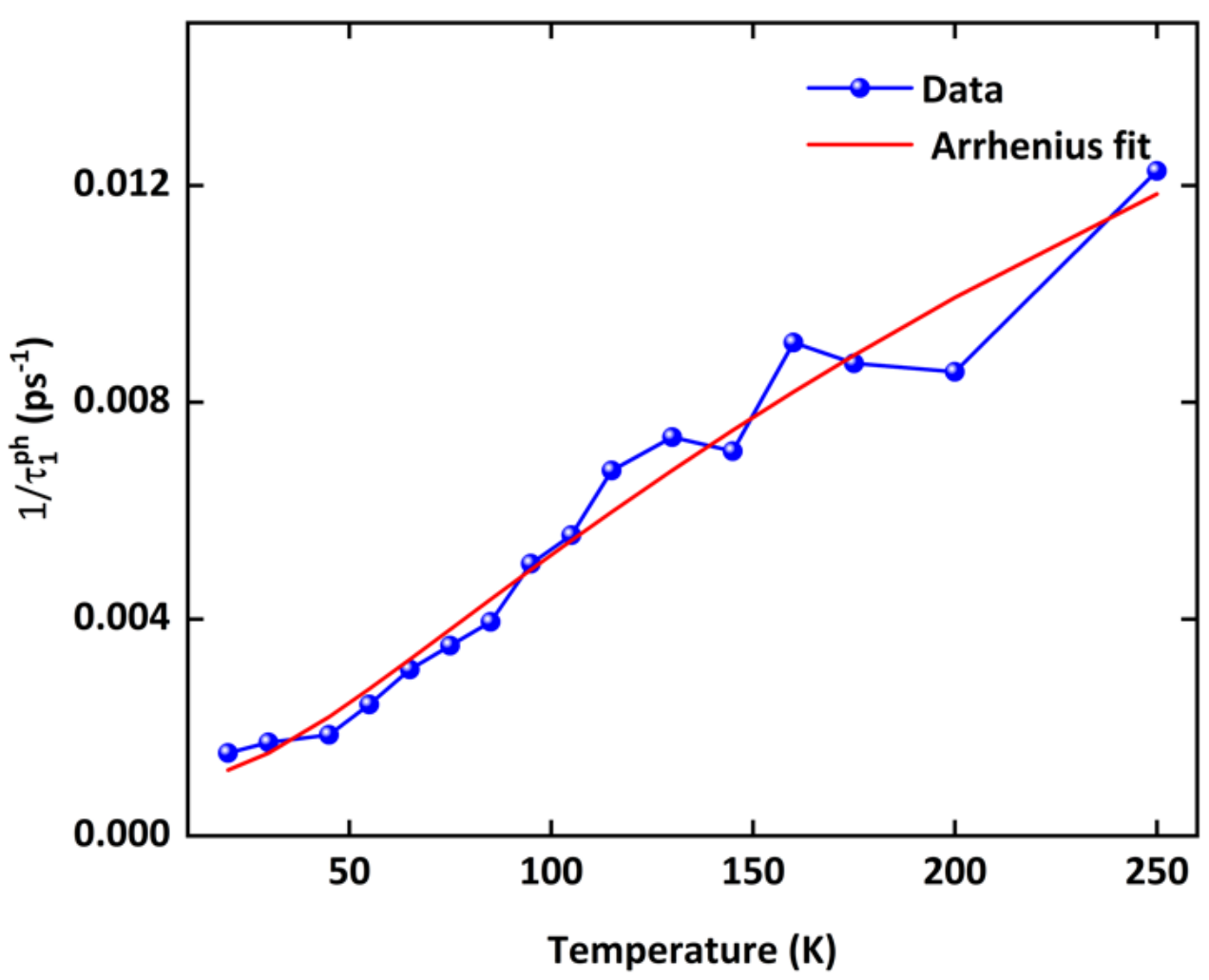

Figure S 13: **Temperature dependence of the phonon scattering rate, 1/τ, of the 0.61 THz coherent phonon mode**. The inverse lifetime (blue) is fitted using Eq. S5(red).

The temperature dependence of the inverse phonon lifetime fits the modified Arrhenius model (Fig. S 13) with fitted parameters A = $1.92\text{x}10^{-2}$ ps$^{-1}$, $\frac{1}{\tau_0}$ = $1.5\text{x}10^{-3}$ ps$^{-1}$, and an activation energy of $E_a$ ~ 14 meV. The extracted activation energy is considerably smaller than the excitonic gap of approximately (80 meV) at 5 K and it is temperature independent. A more accurate model would be to estimate the temperature-dependent gap using a phenomenological two-channel decay model.

**b2.**

Here we consider a simple two-channel decay model derived from Arrhenius model that relates the enhancement of the coherent phonon lifetime to the suppression of a thermally activated non-radiative decay channel by the opening of an energy gap. The enhancement factor is defined as

$$\eta(T) \equiv \frac{\tau(T)}{\tau(T_c^+)}$$

where τ(T) is the coherent phonon lifetime at temperature T and $\tau(T_C^+)$ is the lifetime just above the transition temperature.

The model assumes that the total phonon decay rate consists of two independent contributions: (i) a temperature-independent intrinsic radiative decay rate, $1/\tau_r$ and (ii) a thermally activated non-radiative decay channel with a maximum rate $1/\tau_{nr,0}$. Below the transition temperature, the opening of an energy gap Δ(T) suppresses the non-radiative channel according to a Boltzmann factor, giving

$$\frac{1}{\tau(\mathrm{T})} = \frac{1}{\tau_{\mathrm{r}}} + \frac{1}{\tau_{\mathrm{nr,0}}}\exp\left(-\frac{\Delta(\mathrm{T})}{\mathrm{k_B}T}\right) \qquad Eq.S6$$

Normalizing the lifetime to its value above the transition, where Δ ($T_C^+$) =0, yields,

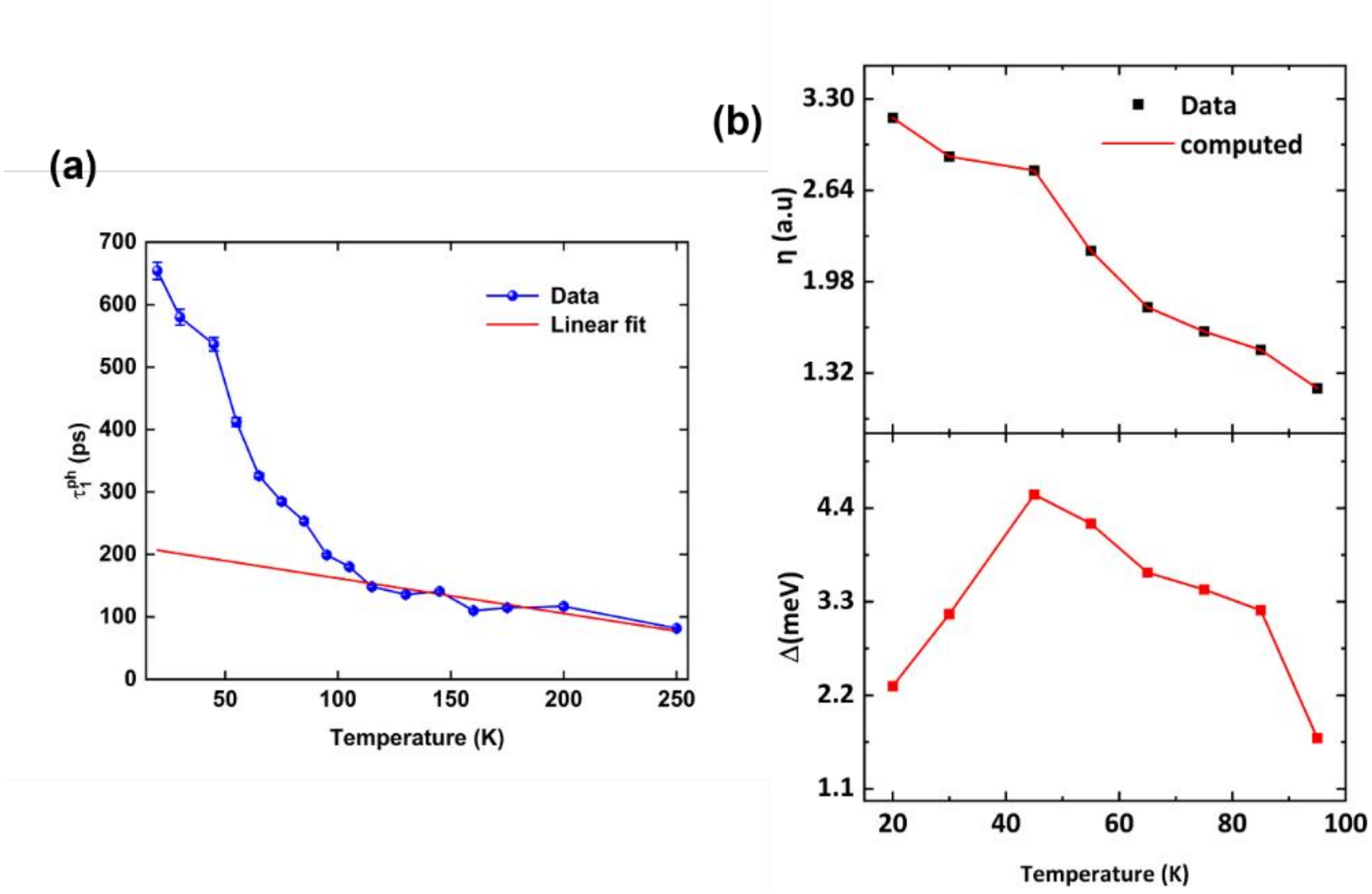

Figure S 14: **Extraction of temperature dependent gap.** (a)Temperature dependence of the lifetime of the 0.61 THz coherent phonon mode. The data (blue) above $T_C$ are fitted with a linear function, which is extrapolated to lower temperatures (red). (b)The lifetime enhancement factor, $\eta$, calculated as the ratio of the measured lifetime to the extrapolated linear fit below $T_C$. The upper panel compares the experimental enhancement factor (black dots) with the values calculated (red) using Eq. (S7), while the lower panel shows the corresponding energy gap extracted from Eq. (S7) as a function of temperature.

$$\eta(T) = \frac{\left[\frac{1}{\tau_r} + \frac{1}{\tau_{nr,0}}\right]}{\left[\frac{1}{\tau_r} + \left(\frac{1}{\tau_{nr,0}}\right)\exp\left(-\frac{\Delta(T)}{k_B T}\right)\right]} \qquad Eq. S7$$

By fitting the phonon lifetime above $T_C$ with a linear function (Fig. S14 a) and extrapolating it to lower temperatures, we estimated the lifetime enhancement below $T_C$ as the ratio $\eta(T)$ (Fig. S14 (b)). Using Eq. (S7), together with the temperature-independent decay rates $1/\tau_r$ and $1/\tau_{nr,0}$obtained from the modified Arrhenius analysis (Eq. S5), we extracted the effective energy gap Δ(T) below $T_C$ as shown in Fig S14 (c). The resulting gap is only a few meV, which is significantly smaller than the excitonic gap reported for $Ta_2Pd_3Te_5$ [5]. This result indicates that opening of the excitonic gap alone cannot account for the observed enhancement of the coherent phonon lifetime, suggesting that additional relaxation mechanisms contribute to the lifetime enhancement below the transition.

### c. Exciton-Polaron model: Lifetime enhancement by polaron formation:

In this section, we explore a possible scenario for the dramatic enhancement of the lifetime of coherent oscillations in our system. The basic idea is that the primary oscillator couples nonlinearly to amplitude fluctuations of the exciton condensate, forming a complex polaron-like state. Since the dressing field is actor ***X*** describes, that the fluctuations of the order parameter of the exciton condensate, we refer to this object as an order-parameter polaron. Here, the role of lattice distortions in the usual polaron picture is taken over by fluctuations of the condensate amplitude.

We model the observed coherent mode as a local oscillator with creation operator $a^\dagger$ coupled to a second local oscillator $b^\dagger$ describing the amplitude fluctuations of the condensate. The minimal independent-boson Holstein Hamiltonian is

$$H_0 = \Omega_a a^\dagger a + \Omega_b b^\dagger b + g a^\dagger a\big(b + b^\dagger\big),$$

where we suppress spatial indices for simplicity. A Lang-Firsov transformation [9] eliminates the linear coupling,

$$U = \exp\big[-\lambda a^\dagger a\big(b^\dagger - b\big)\big],\ \ \lambda = \frac{g}{\Omega_b}$$

and gives, in the single-excitation sector,

$$\tilde{H}_0 = \ \ U\, H_0\, U^\dagger = \Omega_{\mathrm{pol}} a^\dagger a + \Omega_b b^\dagger b,\ \ \Omega_{\mathrm{pol}} = \Omega_a - \lambda^2 \Omega_b.$$

Here, the polaron frequency $\Omega_{pol}$ should be identified with the observed oscillation frequency. The physical annihilation operator is transformed into

$$U a U^\dagger = aX, \qquad X = \exp\big[\lambda\big(b^\dagger - b\big)\big]$$

The extra factor X describes, that the decay of the polaron is associated with a shift of the local exciton amplitude. The relevant finite-temperature correlation function is

$$C_X(t) = \big\langle X(t) X^\dagger(0)\big\rangle_T$$

Using b(t)=b exp (-i$\Omega_b$ t), the displaced-oscillator correlation function becomes

$$C_X(t) = \exp\big\{-\lambda^2\big[(2n_b + 1) - (n_b + 1)e^{-i\Omega_b t} - n_b e^{i\Omega_b t}\big]\big\}$$

$$n_b(T) = \mathrm{n}(\Omega_b, T) = \frac{1}{e^{\beta\Omega_b} - 1}$$

Expanding this periodic function in harmonics,

$$C_X(t) = \sum_m P_m\,(T) e^{-im\Omega_b t}$$

gives the well-known [10, 11] finite-temperature Franck-Condon sideband weights

$$P_m(T) = e^{-\lambda^2(2n_b+1)} e^{\frac{m\beta\Omega_b}{2}} I_m\big(2\lambda^2\sqrt{n_b(n_b+1)}\big) \qquad\qquad Eq.S8$$

where $I_m$ is a modified Bessel function. These weights describe decay events in which *m* quanta are exchanged with the dressing oscillator.

To model the decay of the exciton-dressed polaron, we couple the system to acoustic phonons. In a typical decay process of a q=0 mode, a pair of acoustic phonons with momenta k and -k is generated. We describe this phenomenologically by the finite-temperature bath spectrum $S_B(\omega, T)$

$$S_B(\omega, T) = \omega^4 e^{-\left(\frac{\omega}{\omega_c}\right)^2} \begin{cases} \left[1 + \mathrm{n}\left(\frac{\omega}{2}, T\right)\right]^2, & \omega > 0 \\ \left[\mathrm{n}\left(-\frac{\omega}{2}, T\right)\right]^2, & \omega < 0 \end{cases}$$

With $n(\omega, T) = \frac{1}{\exp\left(\frac{\omega}{T}\right) - 1}$ . The factor $\omega^4$ represents the leading two-acoustic-phonon phase space together with the strain-coupling matrix element, $\omega_c$ is a phenomenological high-frequency cutoff, roughly of the order of the Debye frequency. Positive frequencies describe the emission of two phonons, negative ones their absorption. By construction, $S_B(\omega, T)$ obeys detailed balance,

$S_B(-\omega,T) = e^{-\beta\omega} S_B(\omega,T)$
The golden-rule decay rate is then

$$\Gamma(T) = \Gamma_0 \sum_m P_m(T) S_B\big(\Omega_{\rm pol} - m\Omega_b, T\big)$$

which can be used to model approximately the temperature dependence of the decay rate.

This expression captures several effects. First, dressing the oscillator by large amplitude fluctuations of the exciton condensates suppresses the coupling to acoustic phonons substantially. For low-T this leads to an exponential enhancement of the lifetime by $e^{\lambda^2}$. Due to the $\omega^4$ term, this is also the dominant mode for our parameters. Second, at finite T, the thermal occupation of the excitonic amplitude fluctuations in combination with the thermal occupation of the acoustic phonons give rise to an enhanced decay rate at finite T. The temperature dependence is very sensitive to both the frequency $\Omega_b$ and the effective cutoff frequency $\omega_c$ of the spectral function of the bath.

To model the temperature dependence of the measured lifetime within the exciton-polaron framework, we first fit the experimental lifetime using the anharmonic decay model (Fig. 3(a)) to obtain the reference lifetime, $\tau_0(T)$. The lifetime enhancement factor is then defined as the ratio $\tau(T)/\tau_0(T)$, which isolates the additional enhancement beyond conventional anharmonic phonon decay. This enhancement factor of both the modes is fitted (Fig. S15) using the exciton-polaron model with the polaron frequency fixed at $\Omega_{pol}$ = 29 K for 0.61 THz and $\Omega_{pol}$ = 40 K for 0.85 THz mode, while the dimensionless coupling strength λ, the condensate fluctuation frequency $\Omega_b$, and the acoustic-phonon cut-off frequency $\omega_c$ are treated as free fitting parameters. The obtained fit parameters are summarized in Table S3.

(To fit Fig. 3a in the main text the enhancement factor [$\tau(T)/\tau_0(T)$] is multiplied by the anharmonic lifetime, $\tau_0(T)$, to obtain the lifetime $\tau(T) = \tau_0(T) \times [\tau(T)/\tau_0(T)]$.)

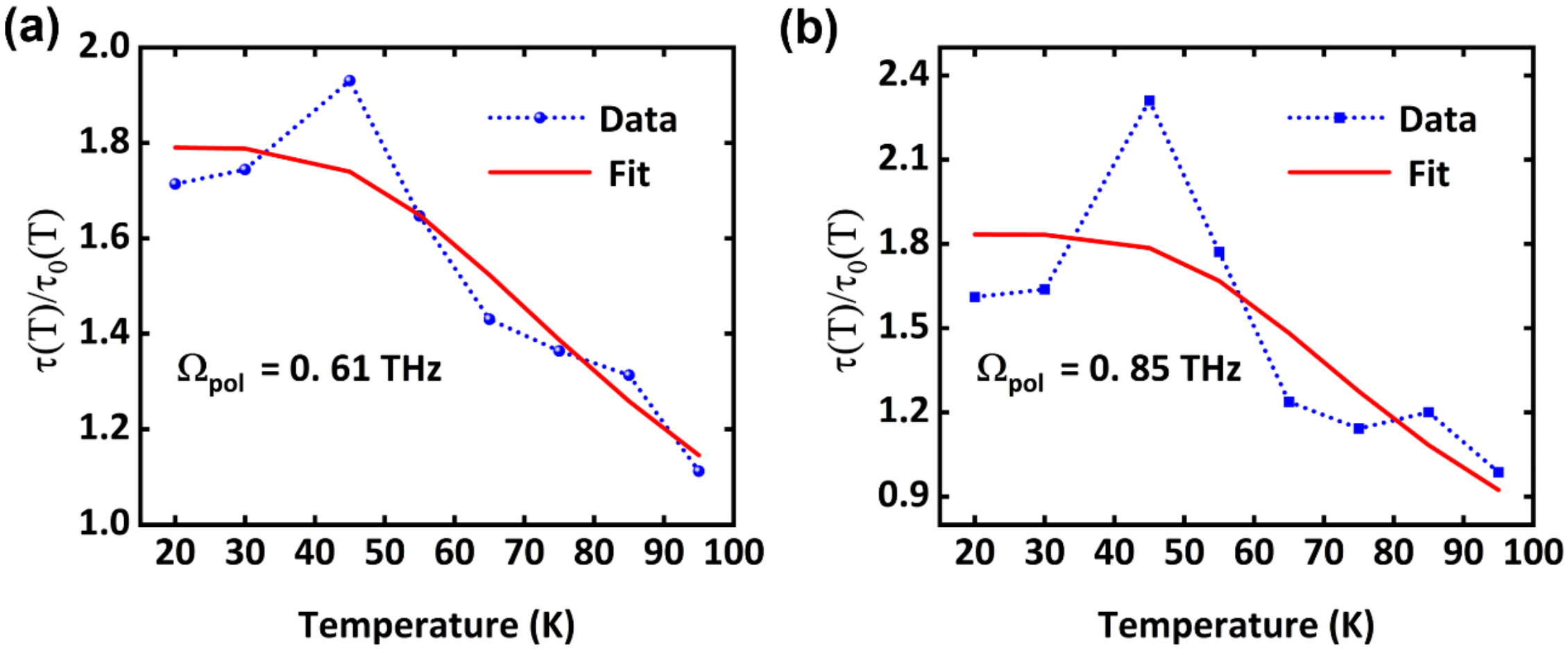

Figure S 15 : **Exciton-Polaron model Fits to the Lifetime Enhancement of the 0.61 and 0.85 THz Modes.** Temperature dependence of the lifetime enhancement factor (blue) together with the fit to the exciton-polaron model (red) for the (a) 0.61 THz mode and (b) 0.85 THz mode.

Table S 3: Fit parameters obtained from the exciton polaron model.

| **Mode frequency (in K)** | **λ** | **$\Omega_b$** | **$\omega_c$** |
|---|---|---|---|
| 0.61 THz (29 K) | 0.76 | 380 ± 90K<br>(32.7meV or 7.9 THz) | 201 ± 64 K<br>(17.3 meV or 4.2THz) |
| 0.85 THz (40 K) | 0.77 | 410 ± 120 K<br>(35.3meV or 8.54 THz) | 212 ± 78 K<br>(18.3 meV or 4.4 THz) |

The coupling constant λ and the characteristic energy scale of the condensate amplitude fluctuations $\Omega_b$ nearly identical for the two coherent phonon modes. This similarity is consistent with both phonons being dressed by the same condensate-amplitude excitation. In addition, the fitted cut-off frequency, $\omega_c$, is of the same order of magnitude as the Debye frequency (~100 K, ≈2 THz) estimated from monolayer ab initio calculations [12], lending further support to the physical plausibility of the model parameters.

Our calculation shows that polaronic dressing can give rise to a massive enhancement of the lifetime of coherent oscillations of low-energy phonon modes. However, we note that the frequency of the coherent oscillations shows very little temperature dependence. This is unlikely in a scenario where a strongly T-dependent amplitude mode is responsible for the polaron formation as the polaron theory predicts a substantial shift of the polaronic energies, $\Omega_{\mathrm{pol}} = \Omega_a - \lambda^2 \Omega_b$. Our highly simplified model neglects several effects, including, for example, non-linear effects of the exciton amplitude mode, temperature dependence of the effective parameters and the dispersion of the modes. Furthermore, it builds on the assumption that the leading decay channel of the order-parameter polaron is acoustic phonons rather than, for example, a disorder-induced mechanism. These effects have to be explored in future works.

### d. Coupled Phonon-Exciton Model (Ph - X): Anharmonic-Phonon Decay with Reduced Electronic Scattering

Here we consider a phenomenological model in which the total damping rate of the coherent phonon contains both a phononic and an electronic contribution. The phononic part is taken from the conventional cubic anharmonic decay process. In addition to this phononic background, we include a phenomenological contribution associated with excitonic gap opening below $T_C$. Since the availability of low-energy electronic states changes strongly upon entering the condensate state due excitonic gap opening, we assume that this damping channel follows the electrical conductivity. Using σ(T) = *1/ρ(T)*, the electronic contribution is written as $\Gamma_{el}$ (T)=B / ρ(T), where B determines the coupling strength of this channel.

The total temperature damping rate is therefore expressed as

$$\Gamma(T) = \Gamma_0 + C\left(1 + \frac{2}{e^{\frac{\hbar\omega_0}{2k_B T}} - 1}\right) + \frac{\mathrm{B}}{\rho(\mathrm{T})}$$

The conductivity-dependent term is motivated by the evolution of the electronic state below $T_C$. As the excitonic gap develops upon cooling, due to suppression in the electronic decay channels the resistivity increases and the conductivity decreases. While this simple heuristic formula is expected to capture the dominant temperature dependence which arises from the density of the electronic excitations, it does not account for the fact that transport and phonon damping probe different electronic vertices and can involve different scattering times. To model the temperature dependence of the phonon lifetime, we used the experimentally measured resistivity $\rho(T)$ (Fig. S2) as an input to the fitting function, while $\Gamma_0$, C and B were treated as free fitting parameters. The experimental data and corresponding fits are shown in Fig. 3(a) for the 0.61 THz mode and in Fig. S10(b) for the 0.85 THz mode, with fit parameters $\Gamma_0 = 8 \times 10^{-4}$ ps$^{-1}$, C = 2.2 x $10^{-4}$ ps$^{-1}$, B = 3.38 x $10^{-5}$ Ω cm ps$^{-1}$ and $\Gamma_0$ = 1.6 x $10^{-3}$ ps$^{-1}$, C = 6.1 x $10^{-4}$ ps$^{-1}$, B = 7.8 x $10^{-6}$ Ω cm ps$^{-1}$, respectively. Note that our model does not explain, why the factor C is apparently much smaller than in other materials.

## 6. Fluence dependence of the 0.61 THz mode

The temperature dependent lifetime and frequency of the 0.61 THz mode for various excitation fluences is shown in Fig. S16. We focus on the prominent non-linear lifetime enhancement regime below the phase transition at $T_C$ = 100 K. As discussed in the main text a slight increase of the fluence reduces this effect significantly and it disappears around 0.38 mJ/cm$^2$ where the lifetime increases linearly on decreasing temperature. This is in line with the picture of the depletion of the condensate in the high fluence regime. Beyond 0.38 mJ/cm$^2$ even a dip around 100K develops.

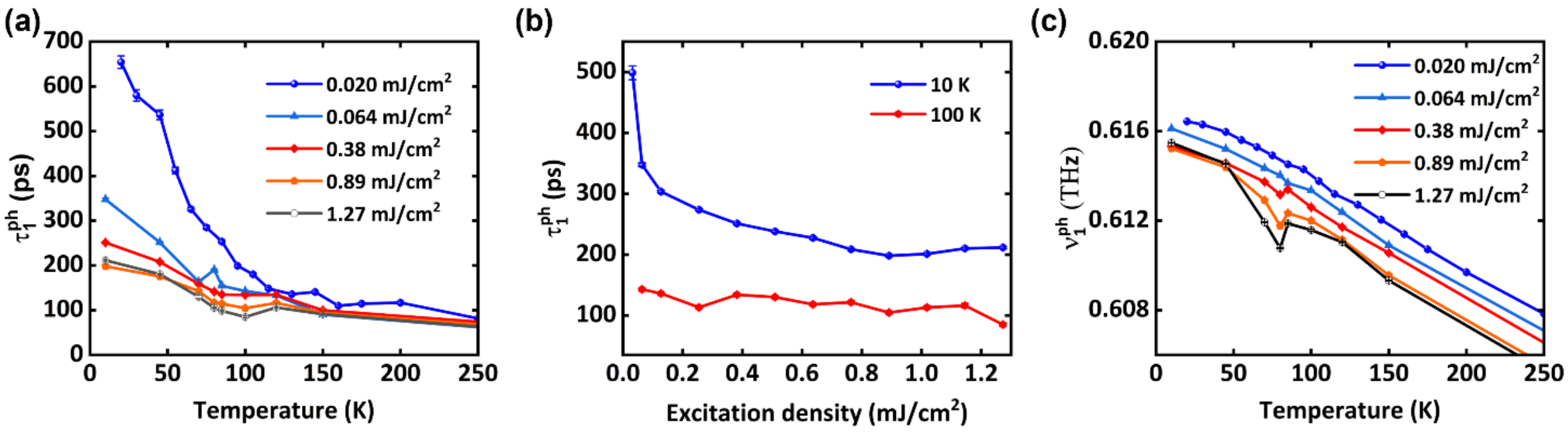

Figure S 16: **Temperature and fluence dependence of the lifetime and frequency of the 0.61 THz mode**. (a) The mode lifetime exhibits a pronounced increase below $T_C$ ≈100 K at low excitation fluence (20 μJ/cm$^2$). At higher excitation fluences above the saturation threshold (~ 0.38 mJ/cm$^2$), a dip /valley-like feature emerges between 80 and 120 K. (b) Fluence dependence of the mode lifetime: in the condensate phase (10 K), the lifetime enhancement undergoes rapid suppression with increasing fluence, whereas at 100 K no lifetime enhancement is present. (c)The temperature dependence of the coupled mode frequency at 0.61 THz shows a deviation from a linear shift as onset of anharmonicities near the transition temperature, $T_C$ ≈ 100 K, at low fluence. With increasing excitation fluence above the saturation threshold, an anomaly in the frequency shift progressively develops.

Figure S16 (b) underlines the fluence dependence of the lifetime at 10 K (blue). A slight increase of the fluence into the saturation regime heavily suppresses the lifetime. In contrast, at the transition temperature of 100 K (red) where no lifetime enhancement takes place, a conventional coherent phonon dynamics of a nearly fluence independent or only slightly linear decreasing lifetime with increasing fluence [13] is observed.

Looking at the frequency dependence for low fluences a deviation from the linear dependence can be detected below 100 K and can be understood as the onset of anharmonicities as discussed in section 5a as shown in Fig. S16(c). The onset of the anharmonicity shifts to lower temperatures on increasing fluence. On fluences beyond 0.38 mJ/cm$^2$ a dip like anomaly around 80 K is observed. This and the corresponding dip in the lifetime enhancement may suggest a photoinduced transition. A strong softening of a collective mode in the excitonic insulator $Ta_2NiSe_5$ was explained as potential light-induced structural transition [14].